\documentclass[%
superscriptaddress,
 amsmath,amssymb,
 aps,
 pra,
floatfix,
]{revtex4-2}
\pdfoutput=1
\usepackage[utf8]{inputenc}
\usepackage[T1]{fontenc}
\usepackage{babel}
\usepackage{graphicx}
\usepackage{amsfonts}
\usepackage{amssymb}
\usepackage{amsmath}
\usepackage{amsthm}
\usepackage{mathtools}
\usepackage{physics}
\usepackage[normalem]{ulem}
\usepackage{blindtext}
\usepackage{dsfont}

\usepackage{braket}
\renewcommand\bra[1]{{\langle{#1}|}}
\renewcommand\ket[1]{{|{#1}\rangle}}
\usepackage{multirow}
\usepackage{tabularx}
\usepackage{dsfont}
\usepackage{bm}
\usepackage{xcolor}

\theoremstyle{definition}

\usepackage{natbib}
\setcitestyle{square,numbers}
\usepackage[colorlinks]{hyperref}
\usepackage[format=plain,justification=centerlast]{caption}
\usepackage[format=plain,justification=centerlast]{subcaption}

\begin{document}

\title{Open dynamics of entanglement in mesoscopic bosonic systems} 

\author{Konrad Schlichtholz}
\email[]{konrad.schlichtholz@phdstud.ug.edu.pl}
\affiliation{International Centre for Theory of Quantum Technologies (ICTQT),
University of Gdansk, 80-308 Gdansk, Poland}
\author{Łukasz Rudnicki}
\affiliation{International Centre for Theory of Quantum Technologies (ICTQT),
University of Gdansk, 80-308 Gdansk, Poland}
\affiliation{Center for Photonics Sciences, University of Eastern Finland, P.O. Box 111, FI-80101 Joensuu, Finland}

\begin{abstract}
A key issue in Quantum Information is finding an adequate description of mesoscopic systems that is simpler than full quantum formalism yet retains crucial information about non-classical phenomena like entanglement. In particular, the study of fully bosonic systems undergoing open evolution is of great importance for the advancement of photonic quantum computing and communication. In this paper, we propose a mesoscopic description of such systems based on boson number correlations. This description allows for tracking Markovian open evolution of entanglement of both non-Gaussian and Gaussian states and their sub-Poissonian statistics. It can be viewed as a generalization of the reduced state of the field formalism [Entropy 2019, 21(7), 705], which by itself does not contain information about entanglement. As our approach adopts the structure of the description of two particles in terms of first quantization, it allows for broad intuitive usage of known tools. Using the proposed formalism, we show the robustness of entanglement against low-temperature damping for four-mode bright squeezed vacuum state and beam-splitted single photon. We also present a generalization of the Mandel Q parameter. Building upon this, we show that the entanglement of the state obtained by beam splitting of a single occupied mode is inherited from sub-Poissonian statistics of the input state.
\end{abstract}

\maketitle

\section{Introduction}
One of the important problems that is still developing in physics is the description of systems on the mesoscopic scale. This problem arises because full quantum treatment of large systems is not feasible. Classical description allows for an accurate depiction of the macroscopic systems; however, one can be interested in systems on the scale on which a full quantum description is not feasible, but with physical phenomena based on quantum features. In such a case, the classical paradigm becomes insufficient, and some intermediate methods are needed for this mesoscopic scale. The need to develop such theoretical tools emerges even more as experiments on this scale become nowadays feasible \cite{Meso_exp1,Meso_exp2}. This need is also fundamental from the perspective of quantum information, since its protocols require complicated quantum systems, and in their principles, they are based on quantum phenomena, such as superposition and entanglement \cite{PPT3}.

Different semi-classical methods were introduced for the simplification of the problems, for example, the widely used parametric approximation \cite{Multiphoton,BGHZ}. However, such techniques still require the full quantum description of some parts of the system. A different approach is to track observables that describe collective behavior of the system, e.g., total spin. Such a mean-field approach captures well macroscopic behavior of the system; however, information about quantum features is suppressed quickly with the number of particles. One can consider another collective observable, namely fluctuations instead of average values, as it is the case for the fluctuation algebra approach to mesoscopic systems \cite{Fluct1,Fluct2}. This technique is adopted in particular to describe entanglement in mesoscopic systems \cite{Fluct1_ent,Fluct2_ent,Fluct3_ent}, and also different phenomena \cite{Carollo_lett,Benatti_2024}. Other methods were also used to describe the entanglement in mesoscopic systems \cite{meso_ent,meso_ent2,meso_reduction}.

One of the specific classes of systems for which mesoscopic descriptions are of great use is fully bosonic systems. The reason for this is that information encoded in bosonic modes is one of the promising avenues for implementing quantum information protocols. This is because, for example, information encoded in photons can be quickly transferred over long distances \cite{Rideout_2012}. However, such systems are also affected by the decoherence effects coming from the environment, and their strict theoretical considerations require dealing with the whole infinite-dimensional Fock space, which can be unfeasible in multi-mode systems. The widely used method that simplifies the description is to track the evolution of the covariance matrix of position and momenta. This reduces the problem to finite-dimensional when a finite number of modes are considered \cite{CV_QI,Isar_2010,Linowski_Cov}. In fact, this approach provides a full description of Gaussian states, and for non-Gaussian states, it can serve as a mesoscopic description. However, non-Gaussian features can be lost in this description, which can result in loss of useful information about quantum features of the state such as entanglement \cite{cumulant_criterion}. Another mesoscopic approach for bosonic systems that was recently proposed is the reduced state of the field (RSF)  \cite{Alicki_Mezo}. This formalism extends the mean-field approach with additional structure by the reduction of the state on the Fock space into the operator on the first quantization single-particle-like Hilbert space.  This description was already applied for the analysis of the coronal heating phenomenon \cite{Alicki_coronal} and the dynamical Casimir effect \cite{Linowski_Rudnicki_2023}. However, it was shown that the RSF approach has some highly classical features and lacks information on quantum features interesting from the standpoint of quantum information. In particular, RSF does not store any information about distillable entanglement \cite{Linowski_Mezo}. Moreover, for two mode Gaussian states, it was shown that it does not contain any information about entanglement.

In this paper, we identify degrees of freedom of bosonic systems that can be used in a mesoscopic description containing information about entanglement of both Gaussian and non-Gaussian states. Upon this, we propose an extended version of the RSF approach, which allows tracking the Markovian evolution of entanglement of bosonic fields. This is achieved by construction of a two-particle-like Hilbert space on which multi-mode bosonic state is mapped into the two-qudit state. Entanglement of this state is sufficient for the entanglement of the full state. In this approach, structures of typical finite-dimensional quantum systems emerge, and therefore one is left with a toolbox which is intuitive and based on well-known tools from quantum mechanics. This approach allowed us to show that the entanglement of the $2\times2$ bright squeezed vacuum state and the entangled two-mode single-photon state is robust against low-temperature damping by the environment. Our approach also allows for tracking particle number variance (fluctuations) and thus for tracking the non-classical phenomenon of sub-Poissonian statistics. In this context, we propose a generalization of the Mandel Q parameter, which can be calculated using the extended RSF. Upon this, we show that the entanglement of the state obtained by beam splitting of one occupied mode is fully inherited from sub-Poissonian statistics of the input state.

\section{Mesoscopic description of bosonic modes}
One of the crucial quantum phenomena is entanglement, which is one of the core resources in many quantum information tasks. This is even more so since in the quantum field theory the entanglement of pure states is fundamentally associated with Bell non-classicality \cite{schlichtholz2023generalization}, which is necessary for device-independent quantum key distribution. For such tasks, particularly compelling is the encoding of information in light modes, as light can be quickly distributed over long distances \cite{Rideout_2012}. What is more, fully photonic quantum devices are one of the promising avenues also in different subfields of quantum information, e.g., quantum computing \cite{PH_Comp}. However, as all quantum systems, light modes are also affected by the environment and undergo non-unitary open evolution. This results in decoherence of the system and thus decay of quantum phenomena such as superposition (including entanglement).

In general, the full description of a non-unitary evolution of the state in terms of infinite-dimensional Fock space can be cumbersome or even infeasible. This is especially true for complicated mesoscopic systems that involve multiple bosonic modes occupied by an undefined number of particles. A custom in the treatment of mesoscopic systems is to consider only some degrees of freedom instead of the full quantum state of the field $\hat\rho_F$. This allows for reducing the complexity of the problem. One of such methods for bosonic systems is to track the evolution of the covariance matrix of position and momenta \cite{Linowski_Cov}:
\begin{equation}
    V_{i,j}=\frac{1}{2}\braket{\{\hat\Xi_i,\hat\Xi_j\}}-\braket{\hat\Xi_i}\braket{\hat\Xi_j},
\end{equation}
where $\vec{\Xi}=(\hat x_1,\hat p_1,....,\hat x_N,\hat p_N)$, and operators $\hat x_j,\hat p_j$ are dimensionless position and momentum operators in $j$-th mode, respectively. This method is particularly important, as it allows tracking of the full Gaussian evolution of Gaussian states. The reason for this is that one can reconstruct the full Gaussian state $\hat\rho_F$  from the covariance matrix. Therefore, the covariance matrix also contains full information about the entanglement of Gaussian states, which in the case of distillable bipartite entanglement can be accessed simply by using the PPT criterion \cite{PPT1,PPT2,PPT3}. However, this is not always the case for non-Gaussian states, e.g., cat-like states. For such states, one has to employ different methods of entanglement detection for which information contained in the covariance matrix is not sufficient \cite{cumulant_criterion}. Furthermore, the covariance matrix approach, while well suited for continuous variable analysis, can be non-intuitive with its symplectic structures. This may particularly concern researchers for researchers working solely with discrete variables, for whom finite-dimensional Hilbert spaces of the first quantization equipped with the Dirac notation can be more familiar. Additionally, regaining information from the covariance matrix about discrete variables also requires extra effort, and it is not always possible for non-Gaussian states.

In fact, multiple fundamental experiments and protocols, e.g., teleportation \cite{Bouwmeester1997}, entanglement swapping \cite{Swapp}, and quantum cryptography \cite{Ekert91} are based on entanglement of qubits realized using light modes restricted to a single photon per party. Emblematic examples of the states used in such protocols are polarization and path-entangled singlet states. Denoting by $\hat a_j^\dagger$ the creation operator in $j$-the mode and vacuum state by $\ket{\Omega}$, such states can be written as: 
\begin{equation}
    \ket{\psi}=\frac{1}{\sqrt{2}}(\hat a^\dagger_1\hat a^\dagger_4-\hat a^\dagger_2\hat a^\dagger_3)\ket{\Omega}=\frac{1}{\sqrt{2}}(\ket{1,0;0,1}-\ket{0,1;1,0}).
\end{equation}
Here, one chooses partition into two parties (subsystems), such that polarization or path modes $a_1,a_2$ are associated with one party and modes $a_3,a_4$ with the second one. Here we have also used the following notation for Fock states:
\begin{equation}
    \ket{n_1,n_2; n_3,n_4}=\frac{(\hat a^\dagger_1)^{n_1}(\hat a^\dagger_2)^{n_2}(\hat a^\dagger_3)^{n_3}(\hat a^\dagger_4)^{n_4}}{\sqrt{n_1!n_2!n_3!n_4!}}\ket{\Omega},
\end{equation}
with semicolon marking the bipartition. Then, the measurements that correspond to Pauli matrices typically considered for qubits are realized as Stokes operators $\hat\Theta_i$ \cite{Tensor} restricted to the single-photon subspace:
\begin{equation}
    \sigma_i\rightarrow\hat\Pi_1\hat \Theta_i\hat\Pi_1= \hat\Pi_1 (\hat a_i^\dagger\hat a_i-\hat a_{i_\perp}^\dagger\hat a_{i_\perp})\hat\Pi_1,
\end{equation}
where $i=1,2,3$ denotes one of the three mutually unbiased basis described by two orthogonal modes $a_i,a_{i\perp}$, and $\hat \Pi_1$ stands for the projector into the subspace of single photon across these modes. Additionally, after projecting with $\hat \Pi_1$, the total photon number operator $\Theta_0=\hat a_i^\dagger\hat a_i+\hat a_{i_\perp}^\dagger\hat a_{i_\perp}$ corresponds to the identity on the qubit space denoted by $\sigma_0$. Note that any local operation on the two-qubit system can be described as a combination of tensor products of Pauli matrices (including $\sigma_0$) acting on subsystems $A$ and $B$. In the Fock space, this translates to $ \sigma_i^A\otimes\sigma_j^B\rightarrow\hat\Pi_1^A\hat\Pi_1^B\hat \Theta_i^A\hat \Theta_j^B\hat\Pi_1^B\hat\Pi_1^A$, where upper indices denote the party on which the given operators act. Therefore, as Stokes operators are, in fact, differences of number operators in two orthogonal modes, all non-classical correlations in such systems are described as photon number correlations $\braket{(\hat a^A_{i(\perp)})^\dagger\hat a^A_{i(\perp)} (\hat a^B_{j(\perp)})^\dagger\hat a^B_{j(\perp)}}$. This suggests that boson number entanglement is at the core of the discrete variable bipartite entanglement of bosonic systems. What is more, such correlations also become experimentally accessible in regimes with higher photon numbers due to developments in the field of photon number resolving detectors \cite{WALMSLEY,WEAKHOMO-OPTICAL-COHERENCE}. Let us note that the considered photonic implementation of qubits is, in fact, approximate. This is because nothing prevents the given modes from being occupied by a higher or lower number of photons. It is an issue even with event-ready preparation of the system in which a photon is confirmed to enter the input of the apparatus, as the environment can add multi-photon noise (with an undefined number of photons) or cause loss of the photon which encodes the information. Such a problem arises especially during propagation through long transmission lines. Thus, even in these highly simplified scenarios, in principle, one should consider the full Fock space, or at least consider some mesoscopic description of the system.  All of this motivates us to build a mesoscopic scale description of bosonic systems with the following matrix as the main structure that contains relevant degrees of freedom of the state $\hat\rho_F$ for considerations of bipartite entanglement:
\begin{align}
        \hat\rho_4:=\sum_{i,j,n,m}^N\tr{\hat\rho_F\hat a_{n}^\dagger\hat a_{i}\hat a_{m}^\dagger\hat a_{j}}\ket{i,j}\bra{n,m}.\label{eq:rho_4}
\end{align}
In the following, we show that $\hat \rho_4$, which we call \textit{extended reduced state}, contains information about entanglement of both Gaussian states with high average numbers of particles, and more importantly, non-Gaussian states (for which the PPT criterion cannot reveal entanglement based on the covariance matrix).  At the same time, we equip the proposed mesoscopic description with multiple intuitive and familiar tools from finite-dimensional Hilbert spaces.

Furthermore, in many circumstances, the open evolution imposed on the system is approximately Markovian. Thus, further on we show the set of evolution equations for the matrix $\hat\rho_4$ for such scenarios.  This type of evolution of the full state of the bosonic modes $\hat\rho_F$ is described by the GKLS equation \cite{GKLS1,GKLS2}. In particular, this equation can be written for $N$ orthogonal bosonic modes $a_j$ using the independent particles approximation in such a way that it includes relevant processes of particle decay and production, random elastic scattering, and classical coherent pumping \cite{Alicki_Mezo,Linowski_Mezo}. We base our considerations upon this equation, which reads:
\begin{align}
\begin{split}
    \frac{d}{dt}\hat\rho_F&=-\frac{i}{\hbar}[\hat H,\rho_F]+\sum_k^{N}[(\xi_k \hat a_k^\dagger-\xi_k^*\hat a_k),\hat\rho_F]+\sum_j\kappa_j(\hat U_j\hat\rho_F\hat U_j^\dagger-\hat\rho_F)\\
    &+\sum_{k,k'=1}^N\Gamma_{\downarrow}^{k',k}\left(\hat a_k\hat \rho_F\hat a_{k'}^\dagger-\frac{1}{2}\{\hat a_{k'}^\dagger\hat a_k,\hat \rho_F\}\right)+\sum_{k,k'=1}^N\Gamma_{\uparrow}^{k',k}\left(\hat a_{k'}^\dagger\hat \rho_F\hat a_{k}-\frac{1}{2}\{\hat  a_{k}\hat a_{k'}^\dagger,\hat \rho_F\}\right),\label{eq:full_ev}
    \end{split}
\end{align}
with Hamiltonian $\hat H= \sum_{k,k'} \omega_{k,k'}\hat a_{k}^\dagger\hat a_{k'}$, where $\hat a_k^\dagger (\hat a_k)$ are creation (annihilation) operators associated with corresponding modes, complex parameters $\xi_k$ describe classical pumping field, $\hat U_j$ are unitary transformations of annihilation operators ($\hat U_j^\dagger\hat a_k\hat U_j=\sum_{k'}u^{j}_{k',k}\hat a_{k'}$ with $u^{j}$ being unitary matrix) corresponding to scattering processes with their probability distribution $\kappa_j$,  and  $\Gamma_\uparrow^{k',k}$ ($\Gamma_\downarrow^{k',k}$) describe creation (annihilation) rate.

\section{Separability in terms of extended reduced state}\label{s:sep}
The key question for entanglement considerations in our mesoscopic description is how separability in the Fock space translates into the extended reduced state. The following considerations are inspired by the mappings of the entanglement indicators for Stokes-like operators \cite{Map,Schlichtholz_Pauli}.
Let us choose two sets of indices $\mathcal{I}_A,\mathcal{I}_B\in \lbrace 1,N\rbrace$ in such a way that $\mathcal{I}_A\cap \mathcal{I}_B=\varnothing$, which determines our bipartition of the system. We assume that the modes with indices from $\mathcal{I}_A$ constitute one subsystem and the modes with indices from $\mathcal{I}_B$  the second one. In addition, each set contains at least two elements, and we denote the number of these elements by $d_X=\overline{\mathcal{I}_X}$. Based on this choice, we project the operator $\hat \rho_4$ in the following way:
\begin{equation}
    \hat \rho_4\rightarrow \hat \rho^\Pi= \hat \Pi_{\mathcal{I}_A,\mathcal{I}_B}\hat \rho_4 \hat\Pi_{\mathcal{I}_A,\mathcal{I}_B}=\sum_{i,n\in \mathcal{I}_A} \sum_{j,m\in \mathcal{I}_B}\rho_{ijnm}\ket{i,j}\bra{n,m},
\end{equation}
where:
\begin{equation}
\Pi_{\mathcal{I}_A,\mathcal{I}_B}=\sum_{i\in \mathcal{I}_A} \sum_{j\in \mathcal{I}_B}   \ket{i,j}\bra{i,j},    
\end{equation}
and $\rho_{ijnm}=\braket{\hat a_{n}^\dagger\hat a_{i}\hat a_{m}^\dagger\hat a_{j}}$ are corresponding matrix elements of $\hat\rho_4$. Let us here for clarity rename the operators corresponding to modes $\mathcal{I}_B$ as $\hat b^\dagger_m$ and $\hat b_j$ , i.e., $\braket{\hat a_{n}^\dagger\hat a_{i}\hat a_{m}^\dagger\hat a_{j}}\rightarrow\braket{\hat a_{n}^\dagger\hat a_{i}\hat b_{m}^\dagger\hat b_{j}}$. Note that the modes that are not used are effectively traced out and modes from  $\mathcal{I}_A$ ($\mathcal{I}_B$) are associated only with the first (second) index of the kets and bras.

Let us assume that the state of the system $\hat \rho_F$ after tracing out modes not contained in $\mathcal{I}_A,\mathcal{I}_B$  is some pure state $\hat\rho_F=\ket{\psi}\bra{\psi}$. In such a case, we can see that $\rho_{ijnm}$ has the form:
\begin{equation}
    \rho_{ijnm}=\langle\psi_{nm}|\psi_{ij}\rangle,
\end{equation}
where $\ket{\psi_{ij}}=\hat a_i\hat a_j\ket{\psi}$. From this it follows that $\rho_{ijnm}$ forms a Gramian matrix and thus is positive definite. Thus $\hat\rho^\Pi_\psi$ (where the subscript denotes the original state on the Fock space) is also positive definite, and after normalization it is a well-defined density matrix $\hat \rho^\mathcal{N}_\psi$ for the system of two qudits with respective dimensions $d_X$. However, there are exceptions for which the matrix cannot be normalized, as it is traceless. The trace of this matrix is equal to $\langle\hat N_{\mathcal{I}_A}\hat N_{\mathcal{I}_B}\rangle$ where $\hat N_{\mathcal{I}_X}$ is the total boson number operator in modes corresponding to $\mathcal{I}_X$. This can be zero only when the state is the vacuum $\hat\rho_F=\ket{\Omega}\bra{\Omega}$ or when in the superposition it consists only of states with bosons only in one subsystem. We address this problem later on. For now, assume that the matrix can be normalized. 
Because any mixed state can be written as a convex combination of pure states $\hat\rho_F=\sum_i p_i\ket{\phi_i}\bra{\phi_i}$, it maps to a convex combination of $\hat\rho^\Pi_{\phi_i}$:
\begin{equation}
    \hat \rho^{\Pi}_{\rho_F}=\sum_i p_i \hat\rho^\Pi_{\phi_i}.
\end{equation}
After normalization, we get:
\begin{equation}
    \hat \rho^\mathcal{N}_{\rho_F}=\sum_i p_i \hat\rho^\Pi_{\phi_i}/\sum_i p_i \tr{\hat\rho^\Pi_{\phi_i}}=\sum_i p_i \tr{\hat\rho^\Pi_{\phi_i}} \hat\rho^\mathcal{N}_{\phi_i}/\sum_i p_i \tr{\hat\rho^\Pi_{\phi_i}}=\sum_i p_i' \hat\rho^\mathcal{N}_{\phi_i}.\label{eq:mix}
\end{equation}
Note that $1\leq p_i'\leq 0$ and $\sum_i p_i'=1$. Thus, this again forms a proper density matrix on the two-qudit space.

In the next step, we assume that $\hat \rho_F$ is separable in the partition into modes $\mathcal{I}_A$ and $\mathcal{I}_B$.  We start with the case  where $\hat\rho_F $ is in some pure state $\ket{\psi}\bra{\psi}$. Due to separability, the state can be written as $\ket{\psi}=F^\dagger_{\mathcal{I}_A}F^\dagger_{\mathcal{I}_B}\ket{\Omega}$, where $\ket{\Omega}$ stands for vacuum state, and $F^\dagger_{\mathcal{I}_i}$ is a polynomial of creation operators of modes corresponding to $\mathcal{I}_X$. From this follows that each $\rho _{ijnm}$ from $\hat\rho^\Pi$ factorizes as follows:
\begin{multline}
    \rho _{ijnm}=\bra{\Omega}F_{\mathcal{I}_A}F_{\mathcal{I}_B}\hat a_n^\dagger\hat a_i\hat b_m^\dagger \hat b_j    F^\dagger_{\mathcal{I}_A}F^\dagger_{\mathcal{I}_B}\ket{\Omega}\\=\bra{\Omega}F_{\mathcal{I}_A}\hat a_n^\dagger\hat a_iF^\dagger_{\mathcal{I}_A}\ket{\Omega} \bra{\Omega}F_{\mathcal{I}_B}\hat b_m^\dagger\hat b_jF^\dagger_{\mathcal{I}_B}\ket{\Omega}=\langle\psi_n^A|\psi_i^A\rangle \langle\psi_m^B|\psi_j^B\rangle, 
\end{multline}
where $\ket{\psi^A_i}=\hat a_i F^\dagger_{\mathcal{I}_A}\ket{\Omega}$ and $\ket{\psi^B_j}=\hat b_j F^\dagger_{\mathcal{I}_B}\ket{\Omega}$. To see this, consider a rearrangement of the creation and annihilation operators in $\rho_{ijnm}$ such that the operators corresponding to $\mathcal{I}_B$  are shifted to the right. Let us then put the resolution of identity $\mathds{1}=\sum_{\vec{n}_1,\vec{n}_2}\ket{\vec{n}_1;\vec{n}_2} \bra{\vec{n}_1;\vec{n}_2}$ between the operators acting on the modes $\mathcal{I}_A$ and $\mathcal{I}_B$. This results in:
\begin{multline}
    \rho _{ijnm}=\bra{\Omega}F_{\mathcal{I}_A}\hat a_n^\dagger\hat a_iF^\dagger_{\mathcal{I}_A}\mathds{1}F_{\mathcal{I}_B}\hat b_m^\dagger \hat b_j    F^\dagger_{\mathcal{I}_B}\ket{\Omega}=\bra{\Omega}F_{\mathcal{I}_A}\hat a_n^\dagger\hat a_iF^\dagger_{\mathcal{I}_A}\ket{\Omega} \bra{\Omega}F_{\mathcal{I}_B}\hat b_m^\dagger\hat b_jF^\dagger_{\mathcal{I}_B}\ket{\Omega}\\+\sum_{\vec{n}_1\neq \vec{0}}\bra{\Omega}F_{\mathcal{I}_A}\hat a_n^\dagger\hat a_iF^\dagger_{\mathcal{I}_A}\ket{\vec{n}_1;\vec{0}} \bra{\vec{n}_1;\vec{0}}F_{\mathcal{I}_B}\hat b_m^\dagger\hat b_jF^\dagger_{\mathcal{I}_B}\ket{\Omega}+
    \sum_{\vec{n}_2\neq \vec{0}}\bra{\Omega}F_{\mathcal{I}_A}\hat a_n^\dagger\hat a_iF^\dagger_{\mathcal{I}_A}\ket{\vec{0};\vec{n}_2} \bra{\vec{0};\vec{n}_2}F_{\mathcal{I}_B}\hat b_m^\dagger\hat b_jF^\dagger_{\mathcal{I}_B}\ket{\Omega}\\+\sum_{\vec{n}_1,\vec{n}_2\neq \vec{0}}\bra{\Omega}F_{\mathcal{I}_A}\hat a_n^\dagger\hat a_iF^\dagger_{\mathcal{I}_A}\ket{\vec{n}_1;\vec{n}_2} \bra{\vec{n}_1;\vec{n}_2}F_{\mathcal{I}_B}\hat b_m^\dagger\hat b_jF^\dagger_{\mathcal{I}_B}\ket{\Omega},\label{eq:sep_d}
\end{multline}
where $\ket{\vec{n}_1;\vec{n}_2}$ stands for the Fock basis state with occupancy of the modes  $\mathcal{I}_A$ and  $\mathcal{I}_B$  described by vectors $\vec{n}_1,\vec{n}_2$, and $\vec{0}$ stands for 0 photons in given modes. This rearrangement is possible as operators acting on orthogonal modes $\mathcal{I}_A$ and  $\mathcal{I}_B$ commute. One can notice that terms of the first sum appearing in (\ref{eq:sep_d}) contain factor $\bra{\vec{n}_1;\vec{0}}F_{\mathcal{I}_B}\hat a_m^\dagger\hat a_jF^\dagger_{\mathcal{I}_B}\ket{\Omega}$ which is equal to $0$. This is because all operators inside the expectation value act only on modes $\mathcal{I}_B$ leaving part of the bra and ket corresponding to modes $\mathcal{I}_A$  orthogonal. Analogously, all other sums that appear in (\ref{eq:sep_d}) are equal to $0$, and the relation is proven. Now,  based on the same arguments as above, $\rho^X_{\psi,l,k}:=\langle\psi_k^X|\psi_l^X\rangle$ forms a proper density matrix after normalization on the one-qudit space. Thus, $\hat \rho^\mathcal{N}_{\psi}=\hat \rho^1_{\psi}\otimes\hat \rho^2_{\psi}/\mathcal{N}$ is a separable state on the two-qudit space. For the mixed separable state, we see from (\ref{eq:mix}) that $\hat \rho^\mathcal{N}_{\rho_F}$ is a convex combination of pure separable states and therefore a proper separable mixed state on the two-qudit space. Therefore, if $\hat \rho^\mathcal{N}_{\rho_F}$ is entangled, then $\hat\rho_F$ is entangled.

Let us now address the scenario in which the matrix for the pure state is not normalizable. This was not important for mapping of entanglement indicators \cite{Map,Schlichtholz_Pauli}.  However, it is important if one wants to use $  \hat \rho^\mathcal{N}_{\rho_F}$  in a more general context as a density operator. All pure states that result in the not normalizable $  \hat \rho^\Pi_{\psi}$ have the following form:
\begin{equation}
    \ket{\psi}=\zeta_0\ket{\Omega}+F^\dagger_{\mathcal{I}_A}\ket{\Omega}+F^\dagger_{\mathcal{I}_B}\ket{\Omega},\label{eq:psi_0}
\end{equation}
where $\zeta_0$ stands for amplitude of vacuum state. These states can be either separable or entangled. One can easily find that for all such states $\hat \rho^\Pi_{\psi}=0$. This tells us that for states of the form (\ref{eq:psi_0}) this state reduction does not allow one to access the information about their entanglement without some additional considerations. What is more, all of them are equivalent to vacuum (which is separable) at the level of reduction to $\hat \rho^\Pi_{\psi}$. Despite that, there is a way to gain access to the entanglement of such states in our description. We discuss this further in examples. Note that such states are effectively highly unlikely to be present in experiments due to the presence of noise, for example, thermal noise. Therefore, the important question is the impact of such states if they are part of the mixed state $\hat \rho_F$. Without loss of generality, assume that only $\ket{\phi_0}$ is of the form \eqref{eq:psi_0}:  
\begin{equation}
    \hat \rho^{\Pi}_{\rho_F}=p_0\hat 0+\sum_{i\geq 1} p_i \hat\rho^\Pi_{\phi_i}\rightarrow\hat \rho^\mathcal{N}_{\rho_F}=\sum_{i\geq 1} p_i \tr{\hat\rho^\Pi_{\phi_i}} \hat\rho^\mathcal{N}_{\phi_i}/\sum_{i\geq 1} p_i \tr{\hat\rho^\Pi_{\phi_i}}=\sum_{i\geq 1} p_i'' \hat\rho^\mathcal{N}_{\phi_i},
\end{equation}
where $p_i''$ fulfill relations for probability. Therefore, even in this case $\hat \rho^\mathcal{N}_{\rho_F}$ is a well-defined density matrix, and clearly the previous discussion on separability holds. In fact, the presence of the states \eqref{eq:psi_0} in the density matrix results only in renormalization of probabilities and thus neither helps to detect entanglement nor prevents it. One can observe that all states of the form:
\begin{equation}
    \hat \rho_F=p_1\sum_j q_j \ket{\psi_j}\bra{\psi_j}+p_2\hat \rho',
\end{equation}
with $\ket{\psi_j}$ having the form \eqref{eq:psi_0} and $q_j, p_i$ being independent probability distributions, are equivalent to the state $\hat \rho'$  in terms of $\hat \rho^\mathcal{N}_{\rho_F}$.

Note that separability of  $ \hat \rho^\Pi_{\psi}$  does not imply that $\hat\rho_F$ is separable, since we use only part of the correlations. In other words, the entanglement of the state $ \hat \rho^\Pi_{\psi}$ on the reduced two-qudit Hilbert space is a sufficient condition for the entanglement of $\hat\rho_F$. Importantly, one has to consider at least 4 modes to reveal such a kind of entanglement, as one has to have both reduced subsystems at least two-dimensional. 

\subsection{Impact of mode transformations}
Note that the choice of the basis in the subsystem defined by $\mathcal{I}_i$ has no impact on the detection of entanglement. This is because the local unitary transformation of the annihilation operators that constitutes a basis transformation translates into the local change of basis in the reduced matrix $\rho^\Pi$. Let us consider a general form of such a transformation $\hat a_i=\sum_k U_{ik}\hat c_k$, where $\hat c_k$ are annihilation operators in some new basis, and $U_{ik}$ determines the transformation between bases. Let us consider a transformation in the first subsystem:
 \begin{equation}
     \rho_{ijnm}^{(a)}=\bra{\psi}\hat a_n^\dagger\hat b_m^\dagger\hat a_i\hat b_j\ket{\psi}=\sum_{k,l}U_{i,k}\bra{\psi}\hat c_l^\dagger\hat b_m^\dagger\hat c_k\hat b_j\ket{\psi}U^*_{l,n}=\sum_{k,l}U_{i,k}\rho_{kjlm}^{(c)}U^*_{l,n},\label{basis_c}
 \end{equation}
 where the superscript in $\rho_{ijnm}^{(a)}$ denotes the basis used. This transformation is simply a basis transformation on the subsystem. Such transformations do not affect the separability of the state.
 
 From this we see that by choice of partition of modes our reduction procedure maps boson number entanglement into the entanglement of two distinguishable particles with finite number of levels. The extended reduced state $\hat\rho_4$ contains information about all partitions $\mathcal{I}_A,\mathcal{I}_B$. Thus, the separability of $\hat\rho_{\rho_F}^\mathcal{N}$ for one partition does not necessarily mean that $\hat\rho_4$ does not contain any information on the entanglement. In fact, $\hat\rho_4$ contains information about the boson number entanglement in any partition $\mathcal{I}_A,\mathcal{I}_B$ after an arbitrary unitary basis transformation of all modes, which can be seen analogously to \eqref{basis_c}. Still, sensible partitions easily emerge in the experimental setups, where specific modes correspond to different beams clearly determining local operations. 

\section{Open evolution and reduced state of the field approach}
In the previous section, we effectively mapped the relevant in the context of entanglement degrees of freedom of a state $\hat\rho_F$ from the infinite-dimensional Fock space into the mixed state $\hat \rho^\mathcal{N}_{\rho_F}$ on two-particle-like finite-dimensional Hilbert space. It is important to note that while $\hat \rho^\mathcal{N}_{\rho_F}$ possesses all mathematical features of the density matrix, it does not inherit a probabilistic interpretation of the entries on the diagonal of the matrix. Here, they describe the strength of linear correlations of boson numbers between given modes. This situation is similar to the case of the reduced state of the field approach (RSF) to the mesoscopic description of bosonic modes \cite{Alicki_Mezo}. In fact, our reduction can be seen as a higher-order extension of the RSF. 

Let us briefly recall the RSF approach.
The RSF formalism attempts to reduce the infinite-dimensional description of the second quantization into a single-particle-like finite-dimensional Hilbert space of the first quantization.
This reduction is carried out as follows. Consider a density operator $\hat \rho_F$ on the Fock space that describes the state of $N$ orthogonal modes $\{a_i\}_{i=1}^N$. This operator is reduced to two structures on the $N$-dimensional Hilbert space equipped with orthonormal basis $\{\ket{k}\}_{k=1}^N$, single-particle density matrix and averaged field:
\begin{equation}\label{eq:rho}
    \hat \rho=\sum_{k,k'=1}^N\tr{\hat\rho_F\hat a_{k'}^\dagger\hat a_k}\ket{k}\bra{k'},\,\,\,\,\ket{\alpha}=\sum_{k=1}^N\tr{\hat\rho_F\hat a_k}\ket{k}.
\end{equation}
The operator $\hat\rho$ contains information on the occupation of modes and coherences, while $\ket{\alpha}$ about the local phases of the field. Note that $\hat\rho$  and $\ket{\alpha}$ after normalization have the properties of a density matrix and a pure state, respectively; however, they do not inherit the probabilistic interpretation normally associated with quantum states. One can see that the diagonal elements of $\hat \rho$ represent the average occupation of modes instead of probabilities.   The reduction of the state is also accompanied by a reduction of observables.  One can reduce a class of additive observables on the Fock space in the following way:
\begin{equation}
    \hat O=\sum_{k,k'=1}^No_{k,k'} a_k^\dagger\hat a_{k'}\rightarrow\hat o=\sum_{k,k'=1}^No_{k,k'}\ket{k}\bra{k'}.\label{eq:ob_RSF}
\end{equation}
This construction preserves the expectation values:
\begin{equation}
    \tr{\hat\rho_F\hat O}=\tr{\hat\rho\hat o}.
\end{equation}
What is important, one can find the time evolution equations for $\hat\rho,\,\ket{\alpha}$  corresponding to the evolution equation \eqref{eq:full_ev}:
\begin{align}
    \begin{split}
        \frac{d}{d t}\hat\rho&=-\frac{i}{\hbar} [\hat h,\hat\rho]+(\ket{\xi}\bra{\alpha}+\ket{\alpha}\bra{\xi})+\sum_j\kappa_j(\hat u_j\hat\rho\hat u_j^\dagger-\hat\rho)+\frac{1}{2}\lbrace\hat \gamma_\uparrow-\hat \gamma_\downarrow^T ,\hat\rho \rbrace+\hat \gamma_\uparrow,\label{eq:ev_rho}
    \end{split}\\
        \begin{split}
        \frac{d}{d t}\ket{\alpha}&=-\frac{i}{\hbar} \hat h\ket{\alpha} +\ket{\xi}+\sum_j\kappa_j(\hat u_j-1)\ket{\alpha}+\frac{1}{2}(\hat \gamma_\uparrow-\hat \gamma_\downarrow^T)\ket{\alpha},\label{eq:ev_alpha}
    \end{split}
\end{align}
where $\hat h$ stands for reduced Hamiltonian (using \eqref{eq:ob_RSF}), and:
\begin{equation}
    \hat \gamma_{\updownarrow}=\sum_{k,k'}\Gamma_{\updownarrow}^{kk'}\ket{k}\bra{k'},\;\;\;\ket{\xi}=\sum_k\xi_k\ket{k},\;\;\;\hat u_j=\sum_{k,k'}u^j_{k,k'}\ket{k}\bra{k'}.
\end{equation}
This formalism is also equipped with notion of entropy:
\begin{equation}
    S[\hat\rho;\ket{\alpha}]=k_B\tr{(\hat\rho^\alpha+1)\log(\hat\rho^\alpha+1)-\hat\rho^\alpha\log(\hat\rho^\alpha)},
\end{equation}
where $k_B$ stands for Boltzmann constant, and $\hat\rho^\alpha=\hat\rho-\ket{\alpha}\bra{\alpha}$.

\subsection{Extended reduced state of the field description}
The RSF formalism, while simple, has drawbacks limiting its applications. Although it is fully based on quantum considerations, it is a highly classical description \cite{Linowski_Mezo}. This is in the sense that RSF does not contain information about non-classical phenomena like distillable entanglement. Therefore, it was conjectured that RSF does not contain any information on entanglement, which was strictly shown for two-mode Gaussian states. What is more, the entropy of this formalism has properties similar to the semi-classical Wehrl entropy rather than the quantum von Neumann  entropy.  However, it turns out that the information contained in RSF is needed to track the evolution of our \textit{reduced two-particle-like state} $\hat \rho^\mathcal{N}_{\rho_F}$. Moreover, this evolution can be put into a framework that extends RSF in a natural way. This extension can be seen as the simplest extension of RSF that adds quantum features to the formalism and is not equivalent to the covariance matrix formalism. 

Let us show how our extended reduced state $\hat\rho_4$ can be incorporated into an RSF-like formalism. States from the single-particle space of RSF contain information about particle numbers in modes. Thus, a natural concept is to introduce a tensor product of two single-particle spaces to track correlations of particle numbers. The extended reduced state $\hat\rho_4$ is then an operator on the two-particle Hilbert space constructed based on two copies of the single-particle space. Note that reduction of state $\hat \rho_F$ does not always have to result in the operator $\hat \rho_4$ being Hermitian. Thus, it does not have to constitute a proper state on this Hilbert space, but still  $\hat \rho^\mathcal{N}_{\rho_F}$ does. We keep the whole structure $\hat \rho_4$ as it  provides additional useful information on quantum features of the state. We will exploit this information later. Moreover, $\hat \rho_4$ is necessary for deriving a closed set of evolution equations that corresponds to (\ref{eq:full_ev}). We build an extended reduced state $\hat \rho_4$ \eqref{eq:rho_4} to resemble a single-particle density matrix $\hat\rho$ \eqref{eq:rho}, i.e., with indices of annihilation operators responsible for labeling rows and indices of creation operators labeling columns. Analogously to \eqref{eq:ob_RSF}, one can reduce the observables:
\begin{equation}
\hat O_4=\sum_{k_1,k_2,k_3,k_4}o_{k_1,k_2,k_3,k_4} \hat a_{k_1}^\dagger\hat a_{k_2}\hat a_{k_3}^\dagger\hat a_{k_4}\rightarrow \hat o_4=\sum_{k_1,k_2,k_3,k_4}o_{k_1,k_2,k_3,k_4}\ket{k_1,k_3}\bra{k_2,k_4},
\end{equation}
with $\tr{\hat\rho_F \hat O_4}=\tr{\hat\rho_4 \hat o_4}$. We also complement the reduction of the state with two additional structures, a rank 3 tensor:
\begin{align}
        \hat\beta :=\sum_{k_1,k_2,k_3,k_4}\tr{\hat\rho_F\hat a_{k_1}^\dagger\hat a_{k_2}\hat a_{k_3}}\ket{k_2,k_3}\bra{k_1},
\end{align}
and an additional operator on single-particle space:
\begin{equation}
\hat{r}=\sum_{k,k'}\Tr{\hat\rho_F\hat{a}_{k'}\hat{a}_{k}}\ket{k}\bra{k'}.
\end{equation}
These two structures are introduced solely in order to include coherent pumping in the evolution, which will become apparent from the evolution equations. The evolution of operators on two-particle space also requires tracking the evolution in a single-particle space. Thus, this two-particle reduction has to be used together with the single-particle reduction, and the \textit{total reduced state} consists of $\hat\rho_4,  \hat\beta,\, \hat r, \hat \rho,\ket{\alpha}$. The Markovian evolution equations for operators $\hat\rho_4,\, \hat\beta,\, \hat r$ can be obtained analogously to equations (\ref{eq:ev_rho},\ref{eq:ev_alpha}). This is achieved by multiplying both sides of \eqref{eq:full_ev} with a suitable combination of creation and annihilation operators for a given matrix element and then tracing the resulting expressions. Then, after some simplifications done with the commutation relation $[\hat a_i,\hat a_j^\dagger]=\delta_{ij}$, the set of evolution equations can be put in the following form:
\begin{align}
    \begin{split}
        \frac{d}{d t}\hat\rho_4&=-\frac{i}{\hbar} [\hat h\otimes \mathbf{1}+ \mathbf{1}\otimes \hat h ,\hat\rho_4]
        \\&+\left(\ket{\xi}\hat\beta^\dagger+(\ket{\xi}\hat\beta^\dagger)^{\tau_L}+(\ket{\xi}\bra{\alpha}\otimes\mathbf{1})^{\tau_L}+\hat\beta\bra{\xi}+(\hat\beta\bra{\xi})^{\tau_R}+(\mathbf{1}\otimes\ket{\alpha}\bra{\xi})^{\tau_R}\right)\\
        &+\sum_j\kappa_j(\hat u_j\otimes\hat u_j\hat\rho_4\hat u_j^\dagger\otimes u_j^\dagger-\hat\rho_4)+\frac{1}{2}\lbrace(\mathbf{1}\otimes\hat\gamma_\uparrow+\hat\gamma_\uparrow\otimes\mathbf{1})- (\mathbf{1}\otimes\hat\gamma_\downarrow^T+\hat\gamma_\downarrow^T\otimes\mathbf{1}) ,\hat\rho_4 \rbrace\\&+(\hat\rho\otimes\gamma_\downarrow^T)^{\tau_L}+(\gamma_\uparrow\otimes\hat\rho)^{\tau_L}+\gamma_\uparrow\otimes\hat\rho+\hat\rho\otimes\gamma_\uparrow+(\gamma_\uparrow\otimes\mathbf{1})^{\tau_L},\label{eq:ev_rho_4}
    \end{split}\\
       \begin{split}
        \frac{d}{d t}\hat \beta&=-\frac{i}{\hbar}[\hat h\otimes\mathbf{1} ,\hat \beta]-\frac{i}{\hbar}\mathbf{1}\otimes\hat h\hat \beta
        \\&+\left(\hat \rho\otimes\ket{\xi}+(\hat \rho\otimes\ket{\xi})^{\tau_L}-\left[(\hat r\otimes\bra{\xi})^{\tau_R}\right]^{T_2} \right)+\sum_j\kappa_j(\hat u_j\otimes\hat u_j\hat\beta\hat u_j^\dagger-\hat\beta)\\&+\frac{1}{2}(\lbrace\hat\gamma_\uparrow\otimes\mathbf{1}-\hat\gamma_\downarrow^T\otimes\mathbf{1},\hat \beta\rbrace+(\mathbf{1}\otimes\hat\gamma_\uparrow-\mathbf{1}\otimes\hat\gamma_\downarrow^T)\hat\beta)
        +\hat\gamma_\uparrow\otimes\ket{\alpha}+(\hat\gamma_\uparrow\otimes\ket{\alpha})^{\tau_L}\label{eq:ev_b}
        \end{split}\\
        \begin{split}
   \frac{d}{d t}\hat r&=-\frac{i}{\hbar} \hat h \hat r+\ket{\xi}\bra{\alpha}+\frac{1}{2}(\hat \gamma_\uparrow-\hat \gamma_\downarrow^T)\hat r+ T.+\sum_j\kappa_j(\hat u_j\hat r\hat u_j^T-\hat r).  \label{eq:ev_r}
\end{split}
\end{align}
Here, we introduced the following notation: $T_i$ stands for transposition on $i$-th space, $\tau_L$ acts on kets as follows $\ket{n,m}\rightarrow\ket{m,n}$ and $\tau_R$ acts analogously on bras, and ``$T.$,, stands for the transposed expression. Also, for some operator on single-particle space $\hat o$,  we denote $\ket{k_2,k_3}\bra{k_1}\hat o\otimes\mathbf{1}:=\ket{k_2,k_3}\bra{k_1}\hat o$, and $\ket{k_2}\bra{k_3}\otimes\ket{k_1}:=\ket{k_2,k_1}\bra{k_3}$,  and $\ket{k_2,k_3}\bra{k_1}\bra{k_4}:=\ket{k_2,k_3}\bra{k_1,k_4}$. One can observe that if there is no coherent pumping (i.e.,  $\forall_i\,\xi_i=0$), evolution of the operators $\hat \rho_4$ and $\hat \rho$ decouples from evolution of the rest of the operators. Therefore, in such circumstances, $\hat \rho_4$ and $\hat \rho$ can be considered by themselves without considering $\hat \beta$ and $\hat r$, for which there is a lack of a quantum state-like interpretation. Note that the first term of the evolution equation \eqref{eq:ev_rho_4} for $\hat\rho_4$ resembles the Heisenberg equation, and it is responsible for the unitary evolution generated by the Hamiltonian. The remaining terms come from decoherence and pumping terms, as in the case of $\hat\rho$. Let us also comment that $\hat r$ together with $\hat \rho$ and $\ket{\alpha}$ allow for reconstruction of the covariance matrix. Therefore, the extension of RSF by $\hat r$ is the simplest extension of RSF that gives quantum features to this formalism. However, it is trivial in the sense that it reproduces the covariance matrix approach, and it is also not a natural extension of this formalism. At the same time, the extension of RSF by $\rho_4$ follows the (multi)particle-like state interpretation of the original RSF. Thus, when coherent pumping is omitted, it provides the simplest extension of RSF that is natural to this formalism and adds the quantum features to the description. Still, one could make analogously further extensions of RSF using three copies of single-particle space, and so on, to track higher-order correlations in photon numbers.

The important feature of the set of equations (\ref{eq:ev_rho},\ref{eq:ev_alpha}) and (\ref{eq:ev_rho_4}-\ref{eq:ev_r}) is that in fact they represent a finite set of first order in time differential equations. Therefore, if the time dependence of $\hat h,\ket{\xi},   \hat \gamma_{\updownarrow}$ is given by simple functions, one can easily find analytic solutions. Additionally, one can solve this set order by order, as solutions of tensors of lower order are independent of solutions for tensors of higher order. Simply, one solves from the lowest order, i.e., $\ket{\alpha}$ and use the solution as non-homogeneous part for the rest of equations. 

The introduction of implicit linear coupling of different modes in the system Hamiltonian $ \hat H=\sum_{k,k’} \omega_{k,k’} a_k^\dagger\hat a_{k’}$ is equivalent to taking a local approach in deriving the master equation where this coupling is considered to be a perturbation. Although it allows one to work explicitly with modes in desired basis, it is not always valid, and then a global approach should be considered (see \cite{Ricard} for a brief review). In the global approach, one first diagonalizes the system Hamiltonian and then considers the evolution of transformed modes. In the end, one transforms the obtained results by reverse unitary transformation of modes to the original basis. As described further in the text, this can be easily done on the reduced state. Therefore, one can easily apply also global approach. In fact, the unitary transformation that diagonalizes the reduced Hamiltonian $\hat h$ is the same as the unitary transformation of modes that diagonalizes the original Hamiltonian. This allows for consideration of the system solely from the reduced perspective.

\subsection{Local evolution}\label{sec:local}
If one is interested in entanglement in a given bipartition, the structure of interest is the reduced two-particle-like state $\hat \rho^\mathcal{N}_{\rho_F}$. However,  based on \eqref{eq:ev_rho_4}, the evolution of this matrix is, in general, not self-contained, and so the full $\hat \rho_4$ is required. Still, if the evolution is local in terms of a given bipartition, then the evolution of $\hat \rho^\mathcal{N}_{\rho_F}$ is closed. In this case, ``local’’ intuitively means that the evolution does not transfer particles between modes corresponding to two parties determined by the given bipartition or does not damp or pump them in a correlated manner. This translates to $\hat h,\,\hat \gamma_{\updownarrow}$ and all $\hat u_j$ being block diagonal in this bipartition, i.e., they can be written as a direct sum of matrices that act only on degrees of freedom of the given party. This allows the projector $\Pi_{\mathcal{I}_A,\mathcal{I}_B}$ to commute with tensor products of such block diagonal matrices as $\hat h\otimes\mathbf{1}$ in equation \eqref{eq:ev_rho_4}. From this, after projecting \eqref{eq:ev_rho_4} into the subspace corresponding to $\Pi_{\mathcal{I}_A,\mathcal{I}_B}$, we get:
 \begin{align}
    \begin{split}
        \frac{d}{d t}\hat \rho^\Pi_{\rho}&=-\frac{i}{\hbar} [\hat h\otimes \mathbf{1}+ \mathbf{1}\otimes \hat h ,\hat \rho^\Pi_{\rho}]\\
        &+\Pi_{\mathcal{I}_A,\mathcal{I}_B}\left(\ket{\xi}\hat\beta^\dagger+(\ket{\xi}\hat\beta^\dagger)^{\tau_L}+(\ket{\xi}\bra{\alpha}\otimes\mathbf{1})^{\tau_L}+\hat\beta\bra{\xi}+(\hat\beta\bra{\xi})^{\tau_R}+(\mathbf{1}\otimes\ket{\alpha}\bra{\xi})^{\tau_R}\right)\Pi_{\mathcal{I}_A,\mathcal{I}_B}\\
&+\sum_j\kappa_j(\hat u_j\otimes\hat u_j\hat\rho^\Pi_{\rho}\hat u_j^\dagger\otimes u_j^\dagger-\hat\rho^\Pi_{\rho})+\frac{1}{2}\lbrace(\mathbf{1}\otimes\hat\gamma_\uparrow+\hat\gamma_\uparrow\otimes\mathbf{1})- (\mathbf{1}\otimes\hat\gamma_\downarrow^T+\hat\gamma_\downarrow^T\otimes\mathbf{1}) ,\hat \rho^\Pi_{\rho} \rbrace\\
&+\Pi_{\mathcal{I}_A,\mathcal{I}_B}\left[(\hat\rho\otimes\gamma_\downarrow^T)^{\tau_L}+(\gamma_\uparrow\otimes\hat\rho)^{\tau_L}+\gamma_\uparrow\otimes\hat\rho+\hat\rho\otimes\gamma_\uparrow+(\gamma_\uparrow\otimes\mathbf{1})^{\tau_L}\right]\Pi_{\mathcal{I}_A,\mathcal{I}_B},\label{eq:loc_ev}
    \end{split}
    \end{align}
where one simply replaces $\hat \rho_4\rightarrow \hat \rho^\Pi_{\rho}$ in equation \eqref{eq:ev_rho_4} and projects the non-homogeneous part using $\Pi_{\mathcal{I}_A,\mathcal{I}_B}$. This allows restricting the considerations of entanglement in the extended RSF to $\hat \rho^\mathcal{N}_{\rho_F}$ instead of tracking the whole $\hat \rho_4$.

\section{Examples: PPT criterion for multi-mode bosonic fields}
We have shown that the sufficient condition for the entanglement of the full state $\hat\rho_F$ is that the reduced two-particle-like state $\hat \rho^\mathcal{N}_{\rho_F}$ is entangled. However, this does not mean that there exists a state $\hat\rho_F$ such that $\hat \rho^\mathcal{N}_{\rho_F}$ is entangled. Therefore, in the following, we show the existence of such states by presenting two examples. What is more, we show that $\hat \rho^\mathcal{N}_{\rho_F}$ contains information about the entanglement for Gaussian states and for non-Gaussian states for which the PPT criterion for the covariance matrix cannot reveal entanglement.

Because one can apply any entanglement detection method for bipartitie finite-dimensional systems to the $\hat \rho^\mathcal{N}_{\rho_F}$, let us consider one of the most widely known entanglement criteria, i.e., the PPT criterion \cite{PPT3}. This criterion states that it is sufficient if partial transposition of the state results in:
\begin{equation}
    (\hat \rho^\mathcal{N}_{\rho_F})^{T_2}<0 
\end{equation}
for state $\hat \rho^\mathcal{N}_{\rho_F}$ to be entangled.
In the simplest case of a two-qubit density matrix, the PPT criterion is a sufficient and necessary condition for entanglement. Therefore, when $\hat \rho^\mathcal{N}_{\rho_F}$ is a two-qubit density matrix and the PPT criterion does not reveal the entanglement of $\hat \rho^\mathcal{N}_{\rho_F}$, then our reduced two-particle-like state is not able to reveal any entanglement of $\hat\rho_F$. This does not imply that the state $\hat\rho_F$ is necessarily a separable state. This is because the entanglement of $\hat \rho^\mathcal{N}_{\rho_F}$ is only a sufficient condition for the entanglement of $\hat\rho_F$.  In the case of a larger number of modes, if the PPT criterion does not reveal the entanglement of $\hat \rho^\mathcal{N}_{\rho_F}$, the reduced state might still keep information about the entanglement of $\hat\rho_F$. However, in such a case, it has to be accessed using a different criterion. While one needs at least four modes for considerations of entanglement using $\hat \rho^\mathcal{N}_{\rho_F}$, in the following, we show a method of how one can also apply this methodology to reveal the entanglement of two-mode states.

\subsection{Bright squeezed vacuum}
As our first example, we will use the $2\times2$ bright squeezed vacuum (BSV). The non-classicality of this state was considered in stationary scenarios, including the high photon number limit \cite{Ent_BSV_1,Ent_stokes,schlichtholz2021simplified}, and in the context of correlation in Bose-Einstein condensates \cite{BEC}.  Let us consider four orthogonal modes and the bipartition given by  $\mathcal{I}_A=\{1,2\},\,\mathcal{I}_B=\{3,4\}$. The BSV state is obtained by unitary evolution of the vacuum generated by the interaction Hamiltonian \cite{Multiphoton}:
\begin{equation}
    \hat H_{int}=\gamma (\hat a_1^\dagger \hat a_4^\dagger-\hat a_2^\dagger \hat a_3^\dagger)+h.c.,\label{eq:BSV_ham}
\end{equation}
and therefore it is a Gaussian state. Now, the BSV state can be written in the form:
\begin{align}
\begin{split}
\ket{\psi_-}&=
\frac{1}{\cosh^2(\Gamma)}\sum_{n=0}^\infty \frac{\tanh^n(\Gamma)}{n!} (\hat a_1^\dagger \hat a_4^{\dagger}-\hat a_2^\dagger \hat a_3^{\dagger})^n \ket{\Omega}
=
\frac{1}{\cosh^2(\Gamma)}\sum_{n=0}^\infty\sqrt{n+1}\tanh^n(\Gamma)\ket{\psi^n},\label{eq:bsv}
\end{split}
\end{align}
where the parameter $\Gamma=\gamma t$ denotes the amplification gain, which is related to the power of the coherent source impinged on the nonlinear crystal in the parametric down conversion process, and 
\begin{equation}
\label{eq:psi_n}
\ket{\psi_-^n}=\frac{1}{\sqrt{n+1}}\sum_{m=0}^n(-1)^m\ket{(n-m),m;m,(n-m)}.   
\end{equation}
Calculation of $\hat\rho^{\Pi}_{\psi}$ yields:
 \begin{equation}
 \hat\rho^{\Pi}_{\psi}=\begin{pmatrix}
\rho_{1313} & 0 & 0 & 0 \\
0 & \rho_{1414} & \rho_{1423} & 0 \\
0 & \rho_{2314} & \rho_{2323} & 0 \\
0 & 0 & 0 & \rho_{2424} 
\end{pmatrix},  \label{eq:bsv_red}
 \end{equation}
 where we have restricted this matrix only to the subspace determined by $\Pi_{\mathcal{I}_A,\mathcal{I}_B}$, and
 \begin{align}
     \begin{split}
     \rho_{1313}&=\rho_{2424}=\sinh ^4(\Gamma ),\\
     \rho_{1414}&=\rho_{2323}=\sinh ^2(\Gamma ) \cosh (2 \Gamma ),\\
     \rho_{2314}&=\rho_{1423}=-\sinh ^2(\Gamma ) \cosh ^2(\Gamma ).
     \end{split}
 \end{align}
After normalization and applying  the partial transpose, we find that $(\hat\rho^{\mathcal{N}}_{\psi})^{T_2}$ has two eigenvalues:
 \begin{align}
     \begin{split}
    \lambda_1&=\frac{1}{1-3 \cosh (2 \Gamma )},\\
   \lambda_2&=\lambda_3=\lambda_4=\frac{1}{3-\text{sech}(2 \Gamma )}.
     \end{split}
 \end{align}
The eigenvalue $\lambda_1$ is negative for all $\Gamma>0$. Therefore, we applied the PPT criterion to show that the BSV state is entangled for any $\Gamma>0$ and thus for an arbitrarily high average number of bosons. As a consequence, the two-particle space can keep information about entanglement in mesoscopic scenarios. 
\subsection{Single-photon}
A s the second example, let us also consider the state of a single photon symmetrically beam splitted into modes $a_1,a_3$:

\begin{equation}
    \ket{\Psi}=\frac{1}{\sqrt{2}}(\ket{1;0}+\ket{0;1})=\frac{1}{\sqrt{2}}(\hat a^\dagger_1+\hat a^\dagger_3)\ket{\Omega}.\label{eq:single}
\end{equation}
This state is of particular interest as it is non-Gaussian and one cannot reveal its entanglement through the PPT criterion for the covariance matrix (see \ref{app:covariance}).
Clearly, this is a two-mode state, and as we argued to apply the above entanglement criterion, one needs four modes. However, one can trivially extend the state to four modes by including modes $a_2,a_4$ with zero photons occupying these modes:
 \begin{equation}
    \ket{\Psi}=\frac{1}{\sqrt{2}}(\ket{10;00}+\ket{00;10}).\label{eq:trivial_ext}
\end{equation}
Still, this state has the form \eqref{eq:psi_0}, for which the RSF results in $\hat\rho^{\Pi}_{\Psi}=\hat 0$, and thus does not give access to information about entanglement. One can see that such a trivial extension of the two mode state will always result in $\hat\rho^{\Pi}_{\rho_F}$ not revealing entanglement. This is because for such an extension there can be at most one nonzero matrix element $\langle\hat a_1^\dagger\hat a_1\hat a_3^\dagger\hat a_3\rangle$ that is necessarily positive, and so $\hat\rho^{\Pi}_{\rho_F}$ in such a case is always a PPT state. Note that this can be seen as a manifestation of the fact that one is not able to observe entanglement of two-mode entangled state with simple photon counting without ancillary states acting as a reference frame.  

To avoid reaching the form \eqref{eq:psi_0}, instead of (\ref{eq:trivial_ext}), one can perform a different extension of the state to four modes by, for example, introducing coherent states with real amplitude $\alpha$ in the additional modes:
 \begin{equation}
    \ket{\Psi}=\frac{1}{\sqrt{2}}(\ket{1\alpha;0\alpha}+\ket{0\alpha;1\alpha})=\frac{e^{-\alpha^2}}{\sqrt{2}}(\hat a^\dagger_1+\hat a^\dagger_3)\left(\sum_{n=0}^\infty \frac{\alpha^n \hat a^\dagger_2}{\sqrt{n!}}\right)\left(\sum_{m=0}^\infty \frac{\alpha^m \hat a^\dagger_4}{\sqrt{m!}}\right)\ket{\Omega}.\label{eq:Psi}
\end{equation}
This extension resembles performing a homodyne measurement on the single-photon state. However, for example, the assumption of matching energy was not done at this step, which would be required to perform the necessary interferometry for homodyne measurement. This is also not fully equivalent to performing homodyne measurement as one is interested in measuring the number of photons instead of quadratures. Such schemes are sometimes referred to as weak homodyne measurement \cite{Das_2021,SP,SinglePhotonGPY}. Now, using the same partition  $\mathcal{I}_A=\{1,2\},\,\mathcal{I}_B=\{3,4\}$, we get the following normalized reduced state $\hat\rho^{\mathcal{N}}_{\Psi}$:
 \begin{equation}
 \hat\rho^{\mathcal{N}}_{\Psi}=\frac{1}{\alpha^2+\alpha^4}\begin{pmatrix}
0 & 0 & 0 & 0 \\
0 & \frac{\alpha^2}{2} & \frac{\alpha^2}{2} & 0 \\
0 & \frac{\alpha^2}{2} & \frac{\alpha^2}{2} & 0 \\
0 & 0 & 0 & \alpha^4 
\end{pmatrix}  .
 \end{equation}
After partial transposition, there are three eigenvalues of the resulting matrix:
 \begin{align}
     \begin{split}
    \lambda_1&=\frac{\alpha^2-\sqrt{\alpha^4+1}}{2 \left(\alpha^2+1\right)},\\
   \lambda_2&=\frac{\alpha^2+\sqrt{\alpha^4+1}}{2 \left(\alpha^2+1\right)},\\
   \lambda_3&=\lambda_4=\frac{1}{2 \left(\alpha^2+1\right)},
     \end{split}
 \end{align}
 where the first eigenvalue is negative for any finite $\alpha>0$. Therefore, even for two-mode entangled states, one can retrieve information about entanglement from the extended RSF. Note that the lowest value of $\lambda_1$ is obtained for $\alpha=0$. This is because one needs $\alpha\neq0$ to allow normalization of the matrix, yet coherent states do not add entanglement to the system but rather some form of noise. Thus, optimally, one should have $\alpha\ll 1$ in this scenario.
 
\subsection{Time evolution of entanglement}
Let us apply the rest of the extended RSF toolbox to consider how entanglement behaves under non-unitary evolution resulting from decoherence coming from the environment. We first consider that one distributes a single-photon state $\ket{\Psi}$ between two distant parties that want to perform a weak homodyne measurement locally to reveal the entanglement of the state. This is an interesting example because a device-independent quantum key distribution protocol can be constructed based on such a setup \cite{schlichtholz2023singlephoton}, and maintaining entanglement during transmission is necessary for its success. However, the state during transmission of the photon necessarily undergoes decoherence. Let us assume that during the distribution of the state, the transmission line is affected by the thermal environment with temperature $T$, while the local oscillators (coherent states) are approximately not affected.  In this scenario, the operators $\hat\gamma_\updownarrow$ are diagonal and fulfil the relation:
$\hat\gamma_\uparrow=e^{-\hbar \omega /k_bT}\hat\gamma_\downarrow$,
where $\omega$ stands for the angular frequency of the modes, and $k_b$ is Boltzmann constant. The creation rates are given by:
\begin{equation}
   \Gamma^{kk}_\uparrow=\gamma_\omega N(\omega)(\delta_{k1}+\delta_{k3}), 
\end{equation}
 where $\delta_{nm}$ stands for Kronecker delta, and coefficient $\gamma_\omega$ determines the coupling of the bath to modes with frequency $\omega$, and $N(\omega)=1/(e^{\hbar \omega /k_bT}-1)$ is Planck distribution of the average number of photons in the bath \cite{Deco}. Furthermore, as all modes are separated up to the moment of the measurement, there is also no coupling between modes. Consequently, the Hamiltonian is simply a free Hamiltonian for four modes with the same frequency. Based on our discussion in section \ref{sec:local}, for this scenario, we can consider only the evolution of reduced single- and two-particle states: $\hat\rho,\,\hat \rho^\Pi_{\rho_\Psi}$. Due to the trivial free evolution Hamiltonian, the master equation simplifies to: 
\begin{align}
    \begin{split}
        \frac{d}{d t}\hat\rho&=\frac{1}{2}\lbrace\hat \gamma_\uparrow-\hat \gamma_\downarrow^T ,\hat\rho \rbrace+\hat \gamma_\uparrow,
    \end{split}\\
    \begin{split}
        \frac{d}{d t}\hat \rho^\Pi_{\Psi}&=
        \frac{1}{2}\lbrace(\mathbf{1}\otimes\hat\gamma_\uparrow+\hat\gamma_\uparrow\otimes\mathbf{1})- (\mathbf{1}\otimes\hat\gamma_\downarrow^T+\hat\gamma_\downarrow^T\otimes\mathbf{1}) ,\hat \rho^\Pi_{\Psi} \rbrace\\&+\Pi_{\mathcal{I}_A,\mathcal{I}_B},\left((\hat\rho\otimes\gamma_\downarrow^T)^{\tau_L}+(\gamma_\uparrow\otimes\hat\rho)^{\tau_L}+\gamma_\uparrow\otimes\hat\rho+\hat\rho\otimes\gamma_\uparrow+(\gamma_\uparrow\otimes\mathbf{1})^{\tau_L}\right)\Pi_{\mathcal{I}_A,\mathcal{I}_B}.
    \end{split}
\end{align}
Using initial condition for the reduced state $\hat \rho$:
 \begin{equation}
 \hat\rho(0)=\begin{pmatrix}
\frac{1}{2} & 0 & \frac{1}{2} & 0 \\
0 & \alpha^2 & 0& \alpha^2 \\
\frac{1}{2} & 0 & \frac{1}{2} & 0 \\
0 & \alpha^2 & 0 & \alpha^2 
\end{pmatrix},  
 \end{equation}
 we obtain the following time evolution of the $\hat \rho^\Pi_{\Psi}$:
  \begin{equation}
 \hat\rho^{\Pi}_{\Psi}(t)=\begin{pmatrix}
e^{-2 \gamma_\omega  t} \lambda \left( \lambda+1\right) & 0 & 0 & 0 \\
0 & \frac{1}{2} \alpha^2 e^{-\gamma_\omega  t} \left(2  \lambda+1\right) & \frac{1}{2} \alpha^2
   e^{-\gamma_\omega  t} & 0 \\
0 & \frac{1}{2} \alpha^2
   e^{-\gamma_\omega  t} & \frac{1}{2} \alpha^2 e^{-\gamma_\omega  t} \left(2  \lambda+1\right)& 0 \\
0 & 0 & 0 & \alpha^4 
\end{pmatrix},  
 \end{equation}
 where $\lambda=N(\omega) \left(e^{\gamma_\omega  t}-1\right)$. Now, in order to access information about entanglement, we apply the PPT criterion after normalization. In the result, we find that there is again one eigenvalue that can be negative:
\begin{equation}
    \lambda_-(t)=\frac{1}{2\mathcal{N}} \Big[\alpha^4+\lambda e^{-2 t \gamma _{\omega }}  \left(\lambda+1\right)
   -e^{-5 t \gamma _{\omega }} \sqrt{e^{6 t \gamma _{\omega }} \left(\alpha^8 e^{4 t \gamma _{\omega
   }}-\alpha^4 e^{2 t \gamma _{\omega }} \left(2 \lambda \left(\lambda+1\right)-1\right)+\lambda^2 \left(\lambda+1\right){}^2\right)}\Big],
\end{equation}
where $\mathcal{N}$  stands for the normalization factor of $\hat\rho^{\Pi}_{\Psi}(t)$. 

Let us first consider $\lambda_-(t)$ in the limit $t\rightarrow\infty$. In this case, there are two options:
\begin{align}
    \lim_{t\rightarrow\infty}\lambda_-(t)&=N(\omega)^2/(\alpha^2+N(\omega))^2 \,\,\,\,\text{for}\,\, \alpha^2\geq N(\omega),   \\
   \lim_{t\rightarrow\infty}\lambda_-(t)&=\alpha^4/(\alpha^2+N(\omega))^2 \,\,\,\,\,\,\,\,\,\,\,\,\,\,\text{for}\,\,\, \alpha^2< N(\omega).  
\end{align}
Remarkably, if $\alpha\neq0$ and the bath temperature is $T=0$ ($N(\omega)=0$) in this limit, one has $\lambda_-(t)=0$, which suggests that the extended RSF formalism finds the state entangled for any finite time $t$. To see that this is the case, first note that: 
\begin{equation}
    \lim_{N(\omega)\rightarrow0}\lambda_-(t)=\frac{ \alpha^4-\sqrt{\alpha^8+\alpha^4 e^{-2 t \gamma _{\omega }}}}{2\mathcal{N}}. 
\end{equation}
Clearly, the second term of the nominator has a higher absolute value, as it is a decreasing function of time with the limit $t\rightarrow\infty$  equal to $\alpha^4$. Therefore, the second term determines the sign of the nominator to be negative. Because the denominator is always positive (as it is the sum of expectation values containing only photon number operators that are non-negative), we find that $\lambda_-(t)$ is also negative in this case for any finite $t$. This certifies the entanglement of the state in the whole time range. Still,  for any $T>0$, the limit of $\lambda_-(t)$ as $t\rightarrow\infty$  is positive, and therefore at some finite time the entanglement will no longer be seen within the RSF formalism. Let us stress that this does not have to imply that the state is no longer entangled, as RSF provides only a sufficient criterion. This indicates that there is no detectable entanglement of the particular kind considered by RSF, i.e., photon number entanglement based on linear correlations. Solving $\lambda_-(t)=0$  for $t$, one finds that this criterion does not detect entanglement after the critical time $t_c$:
\begin{equation}
    t_c \gamma_\omega=\log \left(1+\frac{\sqrt{2}-1}{2 N(\omega)}\right),\label{eq:tc}
\end{equation}
which clearly goes to infinity for $N(\omega)\rightarrow0$ and to 0 for $N(\omega)\rightarrow\infty$. Interestingly, $t_c$ is independent of $\alpha$ and therefore $\alpha$ only affects the magnitude of $\lambda_-(t)$  and not its qualitative behavior in terms of its sign. 

\begin{figure}[!t]
    \centering
    \begin{subfigure}[b]{0.49\textwidth}
         \centering
         \includegraphics[width=\textwidth]{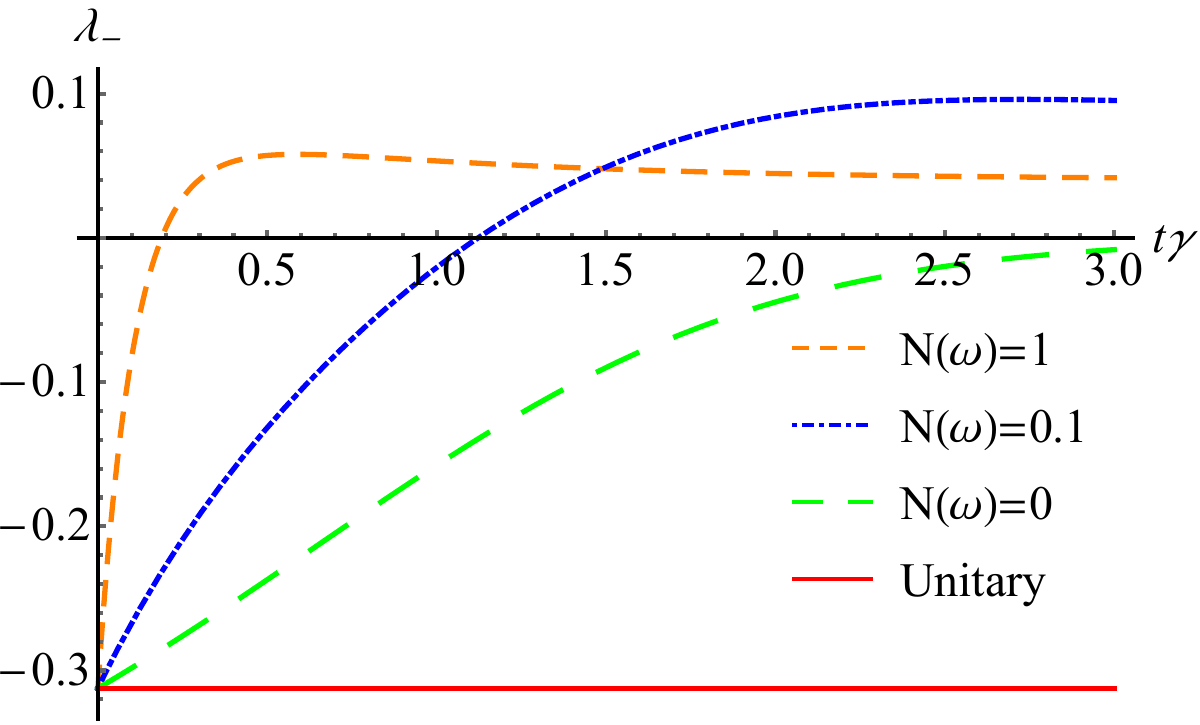}
        \subcaption[]{} \label{fig:1a}
     \end{subfigure} 
      \begin{subfigure}[b]{0.49\textwidth}
         \centering
         \includegraphics[width=\textwidth]{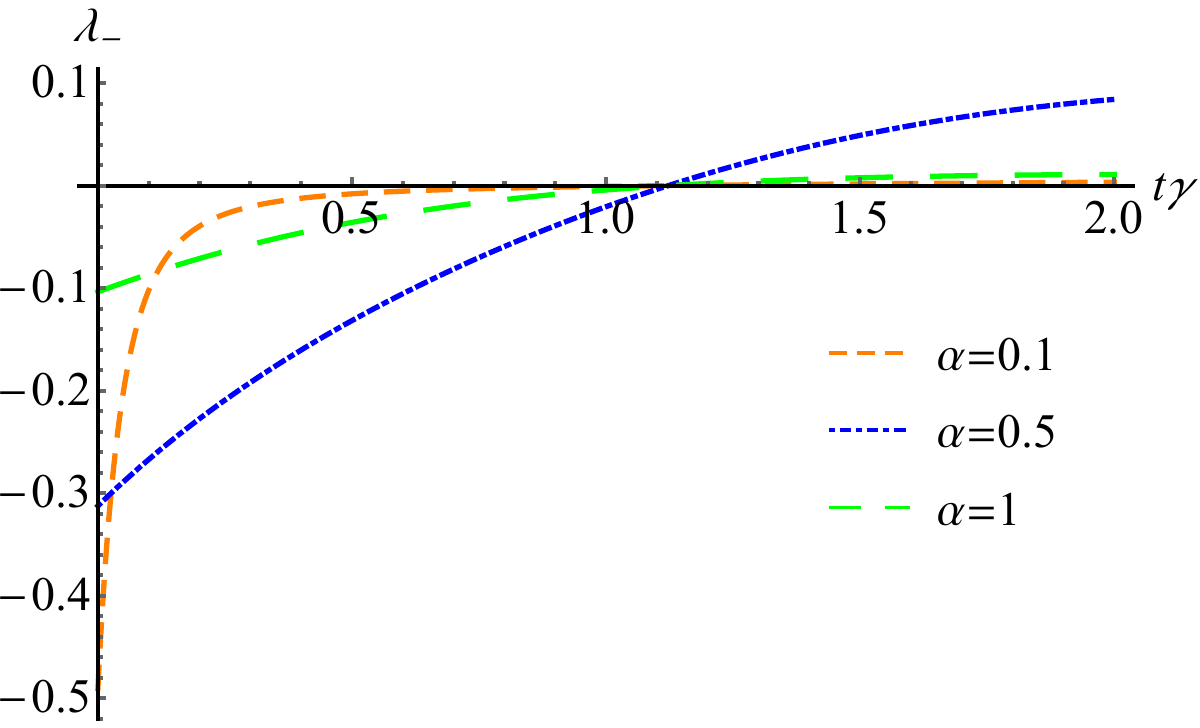}
        \subcaption[]{} \label{fig:1b}
     \end{subfigure}
 
        \caption{a) Time evolution of $\lambda_-(t)$ for $\alpha=0.5$ and different values of $N(\omega)$. The solid red line presents the value of $\lambda_-(t)$ in the case of unitary evolution without a thermal environment. For $N(\omega)=0$  value of $\lambda_-(t)$  converges to zero but is always negative, showing the entanglement of the state.  With increasing $N(\omega)$ the photon number entanglement starts to decay faster, eventually disappearing. b) Time evolution of $\lambda_-(t)$ for $N(\omega)=0.1$ and different values of $\alpha$. For all values of $\alpha$  axis is reached at the same time $t_c$. The value of $\lambda_-(0)$ increases with $\alpha$. However, the rate of growth of $\alpha$ greatly decreases for higher values of $\alpha$. This can be seen as the curve for $\alpha=0.1$  quickly crosses other curves while still being negative, and therefore extremely small values of $\alpha$  are no longer optimal as in the unitary case.   }
        \label{fig:1}  
\end{figure}

Let us comment that, in general, extensions of the state do not have to be equivalent. For example, one can try to engineer an optimal ancilla to maximally violate some Bell inequality \cite{Paterek11}. Here, we also see this kind of inequivalence, since the different extensions (by different coherent states) give different magnitudes of $\lambda_-(t)$. In addition, one can have a completely different behavior, as found, for example, with the trivial extension considered before. Therefore, it is interesting to see that different non-trivial extensions result in the same $t_c$. However, it is not a general property of all non-trivial extensions.  Assuming only that the extended part of the state is separable from the original state, one can find for the considered single-photon state that:
 \begin{equation}
 \hat\rho^{\Pi}_{\Psi}=\begin{pmatrix}
0 & 0 & 0 & 0 \\
0 & \frac{\langle \hat a_4^\dagger \hat a_4\rangle}{2} & \frac{\langle \hat a_2^\dagger \hat a_4\rangle}{2} & 0 \\
0 & \frac{\langle \hat a_4^\dagger \hat a_2\rangle}{2} & \frac{\langle \hat a_2^\dagger \hat a_2\rangle}{2} & 0 \\
0 & 0 & 0 & \langle \hat a_2^\dagger\hat a_2 \hat a_4^\dagger\hat a_4\rangle 
\end{pmatrix}  .
 \end{equation}
From this, one can see that any extension for which $\langle \hat a_2^\dagger \hat a_4\rangle=0$ will not find entanglement. Thus, in this scenario, one excludes any Fock states as meaningful extensions.  Performing calculations analogous to (\ref{eq:tc}), one finds:
\begin{equation}
t_c \gamma_\omega=\log \left(1+\frac{\sqrt{1+|\langle \hat a_2^\dagger \hat a_4\rangle|^2/\langle \hat a_2^\dagger\hat a_2 \hat a_4^\dagger\hat a_4\rangle}-1}{2 N(\omega)}\right),
\end{equation}
which clearly depends on the form of the extension. Then, for the single-photon state (\ref{eq:single}), the optimal extensions are the ones that maximize $|\langle \hat a_2^\dagger \hat a_4\rangle|^2/\langle \hat a_2^\dagger\hat a_2 \hat a_4^\dagger\hat a_4\rangle$. If the ancillary state is a separable state, then by the Cauchy-Schwarz inequality, one can find that this quantity is bounded by 1 \cite{two_mode_ent}. Thus, since the extension by coherent states saturates this bound, it gives an optimal value of $t_c$. 
Fig. \ref{fig:1} presents examples of the evolution of the eigenvalue $\lambda_-(t)$ in different regimes. Clearly, entanglement becomes weaker with time evolution, and its decay becomes faster with increasing $N(\omega)$. In general, this shows that the single-photon entangled state is a resource that is quite robust against thermal damping when the thermal bath has a low temperature.

Let us note that one can propose one extension which is not built from a separable state. In particular, a second copy of the original state can be taken as such an extension. This still indicates that the observed entanglement is due to the entanglement of the original state. However, in such a case, it is rather sensible that the ancillary modes undergo the same evolution as the modes of the original state. This is because the state of the ancillary modes cannot be prepared locally.  Thus, for the considered scenario, one should take the following  creation rates:
\begin{equation}\label{eq:cr_rates_full}
   \Gamma^{kk}_\uparrow=\gamma_\omega N(\omega)\delta_{kk}.
\end{equation}
For this case, $t_c$ is exactly the same as for the case of the extension by coherent states. Thus, this simple extension by the second copy of the state is also optimal in this case.

\begin{figure}
    \centering
    \includegraphics[width=0.5\linewidth]{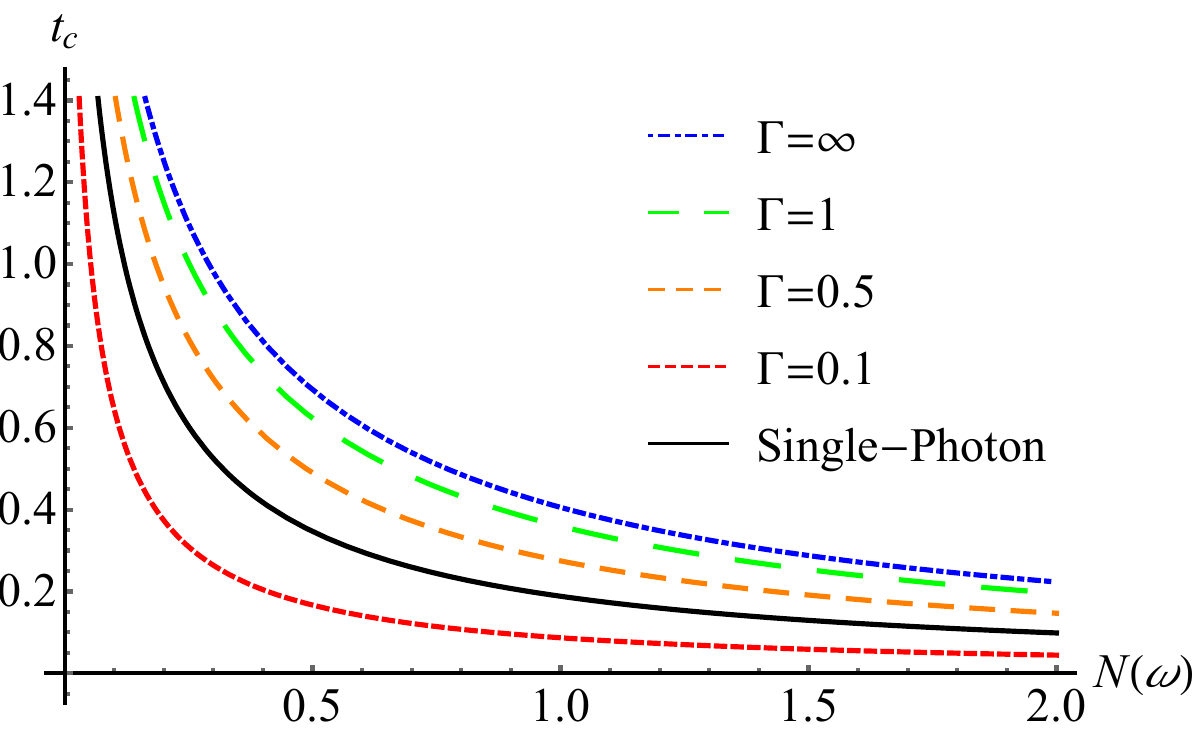}
    \caption{Comparison of $t_c$ for the single-photon state and the BSV state for different $\Gamma$ as a function of $N(\omega)$. The critical time $t_c$  for the BSV state quickly converges to the limiting case $\Gamma\rightarrow\infty$ and reaches a significant advantage in robustness over the single-photon state. Still, for small values of $\Gamma$, entanglement of the single-photon state is more robust against non-unitary evolution resulting from thermal bath.}
    \label{fig2}
\end{figure}

Now, let us consider an analogous scenario for the BSV state where all modes are coupled to the thermal environment with the same temperature during the distribution of the state between parties. Therefore, for this case, we consider the creation rates (\ref{eq:cr_rates_full}).
Then, using as initial conditions reduced states \eqref{eq:bsv_red} and 
\begin{equation}
    \rho=\text{diag}(\sinh^2(\Gamma),\sinh^2(\Gamma),\sinh^2(\Gamma),\sinh^2(\Gamma)),
\end{equation}
we get the following solution for $\hat\rho^{\Pi}_{\psi}$:

    \begin{equation}
 \hat\rho^{\Pi}_{\psi}(t)=\begin{pmatrix}
\rho_{1313}(t) & 0 & 0 & 0 \\
0 & \rho_{1414}(t) & \rho_{1423}(t) & 0 \\
0 & \rho_{2314}(t) & \rho_{2323}(t) & 0 \\
0 & 0 & 0 & \rho_{2424}(t) 
\end{pmatrix},  \label{eq:bsv_red2}
 \end{equation}
where
 \begin{align}
     \begin{split}
     \rho_{1313}&=\rho_{2424}=\frac{1}{4} e^{-2 \gamma_\omega  t} (\cosh (2 \Gamma )+2 \lambda -1)^2,\\
     \rho_{1414}&=\rho_{2323}=e^{-2 \gamma_\omega  t} \left(\cosh (2 \Gamma ) \left(\sinh ^2(\Gamma )+\lambda \right)+(\lambda -1) \lambda \right),\\
     \rho_{2314}&=\rho_{1423}=-e^{-2 \gamma_\omega  t}\sinh ^2(\Gamma ) \cosh ^2(\Gamma ) .
     \end{split}
 \end{align}
 Applying the PPT criterion, we find that the negative eigenvalue evolves as follows: 
 \begin{equation}
     \lambda_1(t)=\frac{1}{4}-\frac{3 \sinh ^2(2 \Gamma )}{2 (8 (2 \lambda -1) \cosh (2 \Gamma )+3 \cosh (4 \Gamma )+16 (\lambda -1) \lambda +5)}.
 \end{equation}
With the realization that the only time dependence is in $\lambda$, which for the special case of $T=0$ is equal to 0, one can see that the entanglement of the BSV state is, in fact, completely stationary under the considered non-unitary evolution, i.e., 
\begin{equation}
    \lambda_1(t)=\lambda_1(0).
\end{equation}
This shows the exceptional robustness of the BSV state to low-temperature damping. For any non-zero temperature ($T\neq0$), there is a point in time at which the entanglement is not detected any more:
\begin{equation}
    t_c \gamma_\omega=\log \left(1+\frac{e^{-\Gamma } \sinh (\Gamma )}{N(\omega)}\right),
\end{equation}
and for $t\rightarrow\infty$ the  eigenvalue $\lambda_1(t)$ converges to $1/4$. Note that $t_c$ is an increasing function of the pumping parameter with the limit 
    \begin{equation}
   \lim_{\Gamma\rightarrow\infty} t_c \gamma_\omega=\log \left(1+\frac{1}{2N(\omega)}\right).\label{eq:tc_bsv}
\end{equation}
Compering this expression to critical time for the single-photon state \eqref{eq:tc}, one can see that the critical time for the highly pumped BSV state is more robust against this type of decoherence. Still, for small values of $\Gamma$ where the BSV state is dominated by the vacuum component, $t_c$ can be smaller than for the single-photon state. This happens for $\Gamma<-\log(2-\sqrt{2})/2\approx 0.27$. Fig. \ref{fig2} shows comparison of $t_c$ for BSV state with different values of $\Gamma$ and single-photon state.

\subsection{Relation with covariance matrix for Gaussian states}
Let us stress that detection of entanglement in a case of some Gaussian states does not imply that $\hat \rho^\mathcal{N}_{\rho_F}$ can reveal entanglement for any entangled Gaussian state.  The reason for this is that while the covariance matrix keeps full information about the Gaussian state, our $\hat \rho^\mathcal{N}_{\rho_F}$ stores information only about a particular type of correlations.

The general inequivalence of these criteria for Gaussian states can be seen based on the two-mode squeezed vacuum state. As we discussed earlier, for the trivial extension of the two-mode state, one is unable to find the entanglement in terms of $\hat \rho^\mathcal{N}_{\rho_F}$, and the proper extension has to be chosen. At the same time, the PPT criterion for the covariance matrix finds that the state is entangled. However, choosing a suitable extension will allow the entanglement of the considered state to be found using $\hat \rho^\mathcal{N}_{\rho_F}$. As in the case of the single-photon state, one can take a second copy of the state as an extension to find the entanglement. Then, the extended state is, in fact, analogous to the BSV state (with the difference that the two ancillary modes are phase-shifted in total by $\pi$), for which $\hat \rho^\mathcal{N}_{\rho_F}$ keeps information about entanglement. To see this, note that the Hamiltonian that generates the BSV state (\ref{eq:BSV_ham}) consists of two commuting parts $\hat H_{14}$ and $\hat H_{23}$ acting on different sets of modes: $\{a_1,a_4\}$ and $\{a_2,a_3\}$. Thus, the unitary evolution of the modes under this Hamiltonian factorizes as $e^{-i\hat H_{14}t/\hbar}e^{-i\hat H_{23}t/\hbar}$  where each exponent corresponds to two-mode squeezing. From this it is apparent that this evolution with the vacuum as the initial state creates a product of two two-mode squeezed vacuum states. However, squeezing parameters for these pairs of modes have the opposite sign based on  (\ref{eq:BSV_ham}).

Observe also that entanglement in Gaussian states is created only through squeezing. In particular, two-mode squeezing creates correlated particle pairs. Therefore, since this type of correlation is in the core of $\hat \rho^\mathcal{N}_{\rho_F}$, it should be efficient in detecting the entanglement of Gaussian states. Moreover, entanglement from single-mode squeezing appears by beam splitting or, more generally speaking, by linear couplings with other modes. From our discussion, we saw that the entanglement created by the beam splitting can be captured by $\hat \rho^\mathcal{N}_{\rho_F}$, and thus is similar in nature.  Consider, as an example, a specific class of entangled Gaussian states. In particular, assume that one starts with the state resulting from some squeezing of different pairs of modes. Then, allow the full state of these modes to evolve based on some Hamiltonian that admits the RSF reduction and couples only modes local to a given party. Before time evolution, one should be able to find if such a state is entangled using $\hat \rho^\mathcal{N}_{\rho_F}$ based on our disscusion (this may require taking a copy of the state as the extension). Now, the evolution equation (\ref{eq:loc_ev})  in our scenario is simply a Heisenberg equation. It generates a unitary evolution that factorizes with respect to subsystems: $e^{-i \hat h_A t/\hbar}\otimes e^{-i\hat h_B t/\hbar}$, where $\hat h_{A(B)}$ is a reduced Hamiltonian restricted to the modes of subsystem A (B). Now, one can write $\hat \rho^\mathcal{N}_{\rho_F}$ in the product basis that diagonalizes $\hat h_{A}$ and $\hat h_{B}$. Importantly, the choice of the particular local basis does not affect whether the PPT criterion finds entanglement. Then, one is left with local free evolution that does not destroy entanglement in the given bipartition. Thus, $\hat \rho^\mathcal{N}_{\rho_F}$ detects entanglement for the considered class of Gaussian states. 

\subsection{Maximally mixed states}
Let us consider the limit $t\rightarrow\infty$ of the reduced two-particle-like state $\hat\rho^{\Pi}_{\psi}(t)$ in \eqref{eq:bsv_red2}. In this limit, one gets:
\begin{equation}
    \lim_{t\rightarrow\infty}\hat\rho^{\Pi}_{\psi}(t)=\text{diag}(N(\omega)^2,N(\omega)^2,N(\omega)^2,N(\omega)^2).
\end{equation}
After normalization, this matrix corresponds to the maximally mixed state on the subspace of the two-particle space restricted to the given bipartition. As the bath in this example was thermal and the same for all modes, this reduced two-particle-like state corresponds to the four-mode thermal state for a given temperature $T$. Note that the temperature does not affect the normalized reduced state. Analogously, we have $\hat \rho=\text{diag}(N(\omega),N(\omega),N(\omega),N(\omega))$, which constitutes a maximally mixed state in a single-particle space. As the inclusion of additional separable modes in the thermal state for a given $T$ cannot add any new type of correlations seen in the reduced states, one can see that the thermal states always correspond to maximally mixed states on single and two particle spaces.

\section{Passive optical elements}
Most optical devices include passive optical elements, such as beam splitters and phase shifters. The RSF formalism allows for easy introduction of such elements into the system. Therefore, this formalism allows for a straightforward analysis of experimental setups. Let us start with the phase shifter. We assume that the beam contains the mode $a_i$ at some point in time, starts the interaction with the phase shifter at $t=\tau_0$ and ends it at $t=\tau_e$, which adds the phase to this particular mode. This can be achieved by the following modification of the Hamiltonian:
\begin{equation}
  \hat H\rightarrow \hat H +\phi_i I_{[\tau_0,\tau_e]}(t)\hat a_i^\dagger \hat a_i,  \label{eq:phase}
\end{equation}
 where $I_{[\tau_0,\tau_e]}(t)$ stands for the indicator function on the time interval $[\tau_0,\tau_e]$, and $\phi_i$ determines the phase gain rate of $i$-th mode. The resulting final phase shift is given by:
 \begin{equation}
     \Delta\phi=(\tau_e-\tau_0)\phi_i.
 \end{equation}
 
The beam splitting of two modes $a_i,\, a_j$ into two modes $a_i',\, a_j'$ can be effectively realized as a unitary transformation of the corresponding annihilation and creation operators given by:

 \begin{equation}
\left( \begin{array}{c}
\hat a_i' \\
\hat a_j '
\end{array} \right) =
\left( \begin{array}{cc}
\sqrt{T} & i\sqrt{R} \\
i\sqrt{R} &\sqrt{T}
\end{array} \right)
\left( \begin{array}{c}
\hat a_{i} \\
\hat a_{j} 
\end{array} \right) =\mathbf{U}_T\left( \begin{array}{c}
\hat a_{i} \\
\hat a_{j} 
\end{array} \right) ,
\end{equation}
 where $R$ and $T$ stand for reflectivity and transmissivity, respectively. Before beam splitting, modes $a_i',\, a_j'$ are in the ground state, and after this transformation, modes $a_i,\, a_j$ are left empty. Therefore, their roles interchange, and tracking modes $a_i,\, a_j$ is of no use after the operation. Therefore, just after the operation, one can rename the modes $a_i',\, a_j'$ to $a_i,\, a_j$, keeping the structure of the reduced state. Based on equation \eqref{basis_c}, one can find that any unitary transformation $\mathbf{U}$  of annihilation operators reduces to the operator:
 \begin{equation}\label{eq:u_tr}
     \mathbf{U}\rightarrow\hat u=\sum_{i,j}U_{ji}\ket{j}\bra{i},
 \end{equation}
where $U_{ji}$ stand for matrix elements of $\mathbf{U}$.  Then, under beam splitting operation, the full reduced state transforms as follows:
 \begin{align}
     \begin{split}
         \hat\rho_4&\rightarrow \hat u_T\otimes\hat u_T\hat\rho_4\hat  u_T^\dagger\otimes\hat u_T^\dagger,\\
         \hat \beta&\rightarrow\hat u_T\otimes\hat u_T\hat\beta\hat  u_T^\dagger,\\
       \hat \rho&\rightarrow \hat u_T\hat\rho\hat u_T^\dagger,\\
        \hat r&\rightarrow\hat u_T\hat r\hat u_T^T,\\
        \ket{\alpha}&\rightarrow\hat u_T\ket{\alpha}.\label{eq:beamsplitt}
     \end{split}
 \end{align}
 Note that one can analogously incorporate to the analysis phase shifting in an instantaneous manner as it results in unitary mode transformation. This reduction of unitary transformations of modes also allows one to consider the inefficiency of the detectors, which can be modeled as a two-port beamsplitter on the path of the mode $a_i$  with transmissivity equal to efficiency $T=\eta$ and second output mode traced out. Denoting the efficiency of the detector at mode $a_i$ as $\eta_i$, inefficient detectors can be applied by operation:
 \begin{equation}
     \hat\eta=\sum_{i}\sqrt{\eta_i}\ket{i}\bra{i}.
 \end{equation}
 Then, the reduced state $\hat \rho$ transforms as $\hat \rho\rightarrow\hat \eta\rho\hat\eta$, and other parts of the reduced state transform analogously.  
 
 Assuming equal efficiency $\eta$ of all detectors, one can see that entanglement of the BSV state can be revealed for arbitrary efficiency, as the matrix $\hat \rho_\psi^\Pi$ \eqref{eq:bsv_red} is varied only by constant coefficient $\eta^2$, which is removed after normalization. Thus, after partial transposition, the eigenvalues are not changed. In fact, this applies to any state and any entanglement criterion by the same means.
 
 Let us go back to the unitary transformation $\mathbf{U}_d$ of the modes that diagonalize the original Hamiltonian $\hat H$. Clearly, the diagonalized Hamiltonian $\hat H_d$ reduces to the diagonal Hamiltonian $\hat h_d$. As the unitary transformation of modes reduces by \eqref{eq:u_tr}, one gets: 
 \begin{equation}\label{eq:diag}
 \hat u_d\hat h\hat u_d^{\dagger}=\hat h_d,
 \end{equation}
 where the equality stands for equality of the matrices representing operators in specific basis in which $h_d$ is diagonal. Therefore, the reduced operators $\hat u_d$ perform the diagonalization of the reduced Hamiltonian. On the other hand, if one finds a unitary operator $\hat u_d$ that fulfills \eqref{eq:diag} for some $\hat h_d$, because the reduction of the Hamiltonian is a bijection, one knows that there is a corresponding diagonal Hamiltonian in the Fock space $\hat H_d$ that must be obtained by mode transformation from the original Hamiltonian $\hat H$. However, since the reduction of unitary transformations of modes is also bijective, this transformation is obtained from $\hat u_d$. 
 \subsection{Homodyne measurement and squeezed states pumping}
 One of the tools used in quantum optics is homodyne measurements. In such measurements on state $\hat\rho_F$, one interferes the coherent state through beamsplitter with state $\hat\rho_F$ and performs intensity measurement or photon counting. Such schemes can be analyzed in the RSF formalism as it requires only extending the reduced state by modes in coherent states (see the \ref{app:extensions}) and performing a beam splitting operation on the obtained reduced state. However, as presented above, in some circumstances one does not even have to mathematically perform a beam splitting operation to get access to information about entanglement.
 
 One of the drawbacks of the RSF formalism is the absence of the possibility of direct inclusion of the squeezing in the evolution of the system. However, there is a partial solution to this problem. In many experimental setups, first some modes are impulsed pumped with squeezed states coming, for example, from parametric down conversion and then combined with the rest of the system through beamsplitters. As one can easily calculate the reduced state for the squeezed states, analogously to the case of homodyne measurements, one can combine the modes in squeezed states with the reduced state of the system. Therefore, one can effectively include some form of squeezing in the system. What is more, one can also include in such a way higher-order squeezed states \cite{BGHZ}.

\section{Mandel-Q parameter and two mode entanglement} 
Our extension of RSF is not limited only to revealing entanglement in terms of non-classical phenomena.  One of the non-classical phenomena in quantum optics is the sub-Poissonian photon statistics of light, which has no classical counterpart. To quantify this phenomenon, one can use the Mandel Q parameter \cite{Mandel}:
\begin{equation}
    Q_i=\frac{\braket{\hat a_i^\dagger\hat a_i\hat a_i^\dagger\hat a_i}-\braket{\hat a_i^\dagger\hat a_i}^2}{\braket{\hat a_i^\dagger\hat a_i}}-1.
\end{equation}
If $Q_i<0$, the state of the mode $a_i$ has sub-Poissonian statistics. This non-classical behavior is still widely studied and used, for example, for source verification \cite{MQ2,MQ1,MQ3}, in particular, using the time-dependent version of this parameter.  Clearly, the extended RSF formalism keeps information on photon statistics and its time dependence in the operators $\hat \rho(t),\hat\rho_4(t)$ and so the Mandel Q parameter can be found as:
\begin{equation}
    Q_i=\frac{\tr\left\{\ket{i,i}\bra{i,i}(\hat\rho_4-\hat\rho\otimes\hat\rho\right)\}}{\tr\{\ket{i}\bra{i}\hat\rho\}}-1.
\end{equation}
This is one of the reasons why one would want to include $\hat \rho_4$ in the extension of RSF instead of only  $\hat\rho^\Pi$. The Mandel Q parameter, which is a shifted and rescaled variance, allows for detecting non-classical behavior internal to the state of a single mode. Still, other parts of the covariance matrix (note that this is a different covariance matrix from the previously mentioned covariance matrix associated with quadratures) that mixes two modes $a_i, \, a_j$ can also be used to reveal non-classicality,  which is shared among multiple modes. Let us build a generalization of the Mandel Q parameter using the criterion for the two-mode entanglement from \cite{two_mode_ent}: 
\begin{equation}
   \braket{\hat a_i^\dagger\hat a_i\hat a_j^\dagger\hat a_j}-\braket{\hat a_i^\dagger\hat a_j}\braket{\hat a_j^\dagger\hat a_i}<0\implies \mathcal{Q}_{ij}\equiv\frac{\braket{\hat a_i^\dagger\hat a_j\hat a_j^\dagger\hat a_i}-\braket{\hat a_i^\dagger\hat a_j}\braket{\hat a_j^\dagger\hat a_i}}{\braket{\hat a_i^\dagger\hat a_i}}-1<0,
\end{equation}
where we have used commutation relation $[\hat a_j,\hat a_j^\dagger]=1$ and the fact that $\braket{\hat a_i^\dagger\hat a_i}$ is non negative. In fact, this parameter contains the covariance of the operators $\hat a_j^\dagger\hat a_i$, $\hat a_i^\dagger\hat a_j$, and clearly it highly resembles the Mandel Q parameter as $\mathcal{Q}_{ii}=Q_i$. In terms of RSF, one can calculate the parameter $\mathcal{Q}_{ij}$ as:
\begin{equation}
    \mathcal{Q}_{ij}=\frac{\tr\left\{\ket{j,i}\bra{i,j}(\hat\rho_4-\hat\rho\otimes\hat\rho\right)\}}{\tr\{\ket{i}\bra{i}\hat\rho\}}-1.
\end{equation}
For the maximally entangled state of a single photon shared among two modes \eqref{eq:single}, one gets $\mathcal{Q}_{13}=\mathcal{Q}_{31}=1/2$. For this state, the time evolution analogous to the one considered in previous sections leads to:
\begin{equation}
    \mathcal{Q}_{13}=\mathcal{Q}_{31}=\frac{(4 \lambda  (\lambda +1)-1) e^{-2 \gamma_\omega  t}}{e^{- \gamma_\omega  t}(\frac{1}{2}-N(\omega))+N(\omega)} .
\end{equation}
One can find that $\mathcal{Q}_{13}=0$  at exactly the same time $t_c$ as the scenario with ancillary coherent states \eqref{eq:tc}.  For $\mathcal{Q}_{13}$,  adding ancillary states to reveal the entanglement turned out to be unnecessary as this parameter uses additional information contained  in $\hat \rho$.

Let us also remark some additional similarities of $\mathcal{Q}_{ij}$ with $Q_i$. Consider $\mathcal{Q}_{12}$ for a two-mode coherent state in modes $a_1,a_2$ with complex amplitudes $\alpha_1,\alpha_2$. In this scenario, one can find that:
\begin{equation}
    \mathcal{Q}_{12}=\frac{|\alpha_1|^2|\alpha_2|^2+|\alpha_1|^2-|\alpha_1|^2|\alpha_2|^2}{|\alpha_1|^2}-1=0.
\end{equation}
Let us recall that whenever $Q_i=0$, the photon counting in the mode $i$ is a Poissonian process, which is always the case for the coherent states. Now, consider a two-mode thermal state:
\begin{equation}
    \hat\rho_{th}=\sum_{i=0}^\infty\frac{N_1(\omega)^i}{(N_1(\omega)+1)^{i+1}}\sum_{j=0}^\infty\frac{N_2(\omega)^j}{(N_2(\omega)+1)^{j+1}}\ket{i;j} \bra{i;j},
\end{equation}
where $N_q(\omega)$ is the average number of photons in mode $q$. For this state, one gets:
\begin{equation}
    \mathcal{Q}_{ij}=N_j(\omega).
\end{equation}
Note that for the thermal state, the Mandel Q parameter has the same value $Q_j=N_j(\omega)$ that exceeds corresponding value in the Poissonian case. One can observe the high resemblance between the values of $\mathcal{Q}_{ij}$ and  $\mathcal{Q}_{ii}$, suggesting a deeper connection between these two.
\subsection{Transformation of sub-Poissonian statistics into two mode entanglement }
Consider two mode state $\hat\rho_{\Phi}$ with one of the modes in an arbitrarily state $\hat\rho_{\varphi}$ and the second one in the zero photon state:
\begin{equation}
    \hat\rho_{\Phi}=\hat\rho_{\varphi}\otimes \ket{0}\bra{0}.\label{eq:prod}
\end{equation}
The relevant reduced states for parameter $\mathcal{Q}_{ij}$ are the following:
\begin{equation}
\hat\rho=\begin{pmatrix}
\braket{\hat n_1} & 0   \\
0 & 0  
\end{pmatrix} ,\,\,\,\,     \hat\rho_4=\begin{pmatrix}
\braket{\hat n_1^2} & 0 & 0 & 0 \\
0 & 0 & 0 & 0 \\
0 & \braket{\hat n_1} & 0 & 0 \\
0 & 0 & 0 & 0
\end{pmatrix}.  \label{eq:RSF_bs}
\end{equation}
Let us now consider a scenario in which one performs the beam splitting operation on this state.  Applying beam splitting transformation \eqref{eq:beamsplitt} to the reduced states (\ref{eq:RSF_bs}), one can calculate parameter $\mathcal{Q}_{ij}^{out}$ for the output modes:
\begin{align}
    \begin{split}
\mathcal{Q}_{12}^{out}&=\frac{\braket{\hat n_1^2}RT+\braket{\hat n_1}T^2-\braket{\hat n_1}^2RT}{\braket{\hat n_1}T}-1,
    \end{split}\\
    \begin{split}
\mathcal{Q}_{21}^{out}&=\frac{\braket{\hat n_1^2}RT+\braket{\hat n_1}R^2-\braket{\hat n_1}^2RT}{\braket{\hat n_1}R}-1,
    \end{split}\\
    \begin{split}
\mathcal{Q}_{11}^{out}&=\frac{\braket{\hat n_1^2}T^2+\braket{\hat n_1}RT-\braket{\hat n_1}^2T^2}{\braket{\hat n_1}T}-1,
    \end{split}\\
  \begin{split}
\mathcal{Q}_{22}^{out}&=\frac{\braket{\hat n_1^2}R^2+\braket{\hat n_1}RT-\braket{\hat n_1}^2R^2}{\braket{\hat n_1}R}-1.
    \end{split}
\end{align}
Note that $\mathcal{Q}_{21}^{out}+\mathcal{Q}_{12}^{out}=\mathcal{Q}_{22}^{out}+\mathcal{Q}_{11}^{out}$,  and so entangled state resulting from the beam splitting has at least one of the modes necessarily sub-Poissonian locally. In fact, both of them have to be sub-Poissonian as:
\begin{align}
    \mathcal{Q}_{11}^{out}&=T\frac{\braket{\hat n_1^2}-\braket{\hat n_1}^2}{\braket{\hat n_1}}+R-1=T\frac{\braket{\hat n_1^2}-\braket{\hat n_1}^2}{\braket{\hat n_1}}-T=T Q_{11}^{in}<0,\\
    \mathcal{Q}_{22}^{out}&=R\frac{\braket{\hat n_1^2}-\braket{\hat n_1}^2}{\braket{\hat n_1}}+T-1=R\frac{\braket{\hat n_1^2}-\braket{\hat n_1}^2}{\braket{\hat n_1}}-R=R Q_{11}^{in}<0,
\end{align}
where we have used the fact that $R+T=1$ and that the input Mandel Q parameter is as follows:
\begin{equation}
    \mathcal{Q}_{11}^{in}=\frac{\braket{\hat n_1^2}-\braket{\hat n_1}^2}{\braket{\hat n_1}}-1.
\end{equation}
Clearly, the sub-Poissonian statistics were redistributed accordingly to transmissivity. Now, it is straightforward to obtain the following particularly important relation:
\begin{equation}
    \mathcal{Q}_{12}^{out}+\mathcal{Q}_{21}^{out}=\mathcal{Q}_{11}^{in}.
\end{equation}
This shows that, starting from the product input state \eqref{eq:prod}, the two-mode entanglement after beam splitting is exclusively inherited from the sub-Poissonian statistics of the input state. Therefore, beam splitting redistributes the internal non-classical photon number correlation of the state to the photon number entanglement of the output modes. Note that due to this equality, $ \mathcal{Q}_{12}^{out}+\mathcal{Q}_{21}^{out}$ is lower bounded by -1 and minimized by the input states being Fock states  $\hat\rho_\varphi=\ket{n}\bra{n}$.
\section{Concluding remarks}
In summary, we have proposed an extension of the reduced state of the field formalism for mesoscopic bosonic fields. Our extension allows for tracking the evolution of particle number entanglement for systems undergoing non-unitary Markovian evolution, which is not possible in the original reduced state of the field approach. The extended approach allows for studying entanglement of both Gaussian and non-Gaussian states. This was achieved by reducing the problem of entanglement of multi-mode bosonic state into the entanglement of the two-particle state on a finite-dimensional Hilbert space. Based on the proposed approach, we showed that particle number entanglement of multi-mode bosonic field is independent of the efficiency of the detectors if it is the same across all detectors. Moreover, we showed that the entanglement of the states of the beam-splitted single photon and $2\times2$ bright squeezed vacuum are robust against interaction with a low-temperature thermal environment. Furthermore, we considered the generalization of the Mandel Q parameter constructed within the proposed approach, which describes non-classical correlations both in a single mode and shared among different modes. Using this parameter, we showed that the entanglement of two-mode entangled states obtained by the beam splitting of one occupied mode is solely inherited from the sub-Poissonian statistics of the input state. 

The proposed reduced state of the field description can provide a versatile and intuitive tool for the analysis of the impact of decoherence on non-classical phenomena in multiple optical experimental setups and consequently their designing. This is because our approach inherits multiple tools from standard quantum mechanics of finite-dimensional systems, with only a slight change in interpretation. As analysis of a given setup requires only specifying mesoscopic properties of the sources instead of the whole initial state, it could be done without even invoking the Fock space. Furthermore, one can easily incorporate passive optical elements and inefficiencies in the description of the evolution. In addition, ancillary modes and entanglement can also be added to the system through the use of beamsplitters.

One of the important limitations of proposed formalism is the class of accessible evolution equations. This class is restricted to master equations with terms that are at most linear in the creation and annihilation operators. Often, it is possible to transform the considered problem into the problem described by such evolution through employing, e. g., Bogoliubov transformations of modes and through simplifying interactions using the mean field. However, if one wants to transform back to the original modes, one has to consider the evolution of additional structures. This is because the information contained in $\rho_4$ is insufficient. Alternatively to Bogoliubov transformations, one can directly calculate the evolution of $\rho_4$ for quadratic Hamiltonians by including the additional structures mentioned above (see \ref{app_second}). However, for higher-order interactions, one needs to track higher-order correlations in order to propose a sensible approximate set of closed evolution equations. Still, whenever one cannot use the convenient evolution equations for the reduced state described in this work, one can use the extended reduced state in the analysis of non-classical features of the state. Another limitation of the formalism is that it is restricted to two-party entanglement and that entanglement of the reduced state is only a sufficient criterion for entanglement. Nevertheless, one can build analogously reduced three- and more-particle-like states and upon it consider more involved types of entanglement.   

Let us also note that since the extended reduced state of the field preserves a state-like interpretation of reduced states and maintains a Hilbert space structure for them, one can exploit scalar product of these Hilbert spaces for geometric-like considerations of bosonic states as, e.g., in \cite{Geo}. This allows for considerations of proximity of states in terms of reduced states and therefore for their classification in terms of mesoscopic properties.

As we found that for both single-photon entangled state and bright squeezed vacuum the entanglement is robust against thermal damping in low temperatures, an interesting open question arises. Is this property universal to photon number entanglement?

The proposed extension of the reduced state of the field approach could also be used to examine different quantum optical phenomena. Let us recall that the original reduced state of the field contained information about the expectation values of Stokes operators, providing a generalization to the description of polarization in terms of Jones vectors. Our extension, when restricted to two modes, additionally keeps information about second-order polarization tensors \cite{Tensor, hidden2}. As these tensors are associated with non-classical phenomena of hidden polarization, our approach could give additional insight into this phenomenon.

\section{Acknowledgements}
The work is part of ‘International Centre for
Theory of Quantum Technologies’ project (contract no. 2018/MAB/5), which is carried out
within the International Research Agendas Programme (IRAP) of the Foundation for Polish
Science (FNP) co-financed by the European Union from the funds of the Smart Growth
Operational Programme, axis IV: Increasing the research potential (Measure 4.3). {\L}R has also been partially supported by the Academy of Finland PROFI funding (336119). The authors thank Tomasz Linowski, Antonio Mandarino, and Agnieszka Schlichtholz for their remarks. 

\bibliography{biblio.bib}
 \appendix
\section{Lack of entanglement detection of beam-splitted single photon by PPT criterion applied to covariance matrix}\label{app:covariance}
Let us show that the PPT criterion applied to the covariance matrix does not allow for the detection of entanglement of the state of single photon beam splitted into two modes $a_1$ and $a_2$:
\begin{equation}
    \ket{\Psi}=\frac{1}{\sqrt{2}}(\ket{1;0}+\ket{0;1})=\frac{1}{\sqrt{2}}(\hat a^\dagger_1+\hat a^\dagger_2)\ket{\Omega}.\label{eq:single_app}
\end{equation}
Let us recall that covariance matrix is given by:
\begin{equation}
    V_{i,j}=\frac{1}{2}\braket{\{\hat\Xi_i,\hat\Xi_j\}}-\braket{\hat\Xi_i}\braket{\hat\Xi_j},
\end{equation}
where $\vec{\Xi}=(\hat x_1,\hat p_1,....,\hat x_n,\hat p_n)$ and operators $\hat x_j,\hat p_j$ are dimensionless position and momentum operators in $j$-th mode:
\begin{equation}
    \hat x_j=\frac{1}{\sqrt{2}}(\hat a_j+\hat a_j^\dagger),\;\;\;\hat p_j=\frac{i}{\sqrt{2}}(\hat a_j^\dagger-\hat a_j).
\end{equation}
By noting that for state \eqref{eq:single_app} we have $\braket{\hat a_j^\dagger}=\braket{\hat a_j}=\braket{\hat a_j^\dagger\hat a_i^\dagger}=\braket{\hat a_j\hat a_i}=0$ and $\braket{\hat a_j^\dagger\hat a_i}=1/2$, one can write down the covariance matrix for this state as:
\begin{equation}
V=\frac{1}{2}\begin{pmatrix}
\braket{2\hat a_1^\dagger\hat a_1}+1 & 0 & \braket{\hat a_1^\dagger\hat a_2+\hat a_2^\dagger\hat a_1} & 0 \\
0 & \braket{2\hat a_1^\dagger\hat a_1}+1 & 0 & \braket{\hat a_1^\dagger\hat a_2+\hat a_2^\dagger\hat a_1} \\
\braket{\hat a_1^\dagger\hat a_2+\hat a_2^\dagger\hat a_1} & 0 & \braket{2\hat a_2^\dagger\hat a_2}+1 & 0 \\
0 & \braket{\hat a_1^\dagger\hat a_2+\hat a_2^\dagger\hat a_1} & 0 & \braket{2\hat a_2^\dagger\hat a_2}+1
\end{pmatrix}=\frac{1}{2}\begin{pmatrix}
2 & 0 & 1 & 0 \\
0 & 2 & 0 & 1 \\
1 & 0 & 2 & 0 \\
0 & 1 & 0 & 2
\end{pmatrix}.  
\end{equation}
The PPT criterion for entanglement in terms of the covariance matrix is given as follows:
\begin{equation}
    Q\cdot V\cdot Q-\frac{i}{2}J<0,\label{eq:PPT_CV}
\end{equation}
where $Q=diag(1,q_1,...,1,q_n)$ with $q_j=-1$ for modes corresponding to transposed subsystem and $q_j=1$ otherwise, and:
\begin{equation}
    J=\bigoplus_{k=1}^n J_2,\,\,\,\,J_2=\begin{pmatrix}
0 & 1  \\
-1 & 0 
\end{pmatrix}.
\end{equation}
This criterion is necessary and sufficient condition for detecting two mode entanglement of Gaussian states and sufficient condition for non-Gaussian states (which is a the case for \eqref{eq:single_app}).  The eigenvalues of the matrix on the r.h.s. of inequality \eqref{eq:PPT_CV} for the state \eqref{eq:single_app} are positive and equal to $(2\pm\sqrt{2})/2$. Therefore, one does not find entanglement with PPT criterion applied to the covariance matrix of the state \eqref{eq:single_app}.
 \section{Extending reduce states by separable subsystems} \label{app:extensions}
 Often when one wants to consider multiple modes with the states of different modes prepared by different sources. In such a case, the total state is a product state of subsystems. Let us consider the case of two subsystems A and B with total state $\rho_{tot}=\hat\rho_A\otimes\hat\rho_B$. It is often easier to construct the reduced state of the total state using the simpler reduced states of $\hat\rho_A$ and $\hat\rho_B$. This is possible as for such states we have $\braket{\hat A\hat B}=\braket{\hat A}\braket{\hat B}$, where $\hat A,\hat B$ are some observables on modes corresponding to subsystems A and B. The reduced state of the whole system can then be obtained as follows:
\begin{align}
    \begin{split}
        \ket{\alpha_{t}}&=\ket{\alpha_{A}}+\ket{\alpha_{B}},\\
        \hat\rho_t&=\hat\rho_A+\hat\rho_B+\ket{\alpha_{A}}\bra{\alpha_{B}}+\ket{\alpha_{B}}\bra{\alpha_{A}},\\
        \hat r_t&=\hat r_A+\hat r_B+\ket{\alpha_{A}}\bra{\alpha_{B}^*}+\ket{\alpha_{B}}\bra{\alpha_{A}^*},\\
    \hat\beta_{t}&=\hat \beta_A +(\ket{\alpha_A^*}\hat r_B)^{T_1}+(\ket{\alpha_B}\hat \rho_A)^{\tau_L}+A\leftrightarrow B,\\
    \hat\rho_{4t}&=\hat\rho_{4A}+\hat\beta_A\bra{\alpha_B}+(\hat\beta_A\bra{\alpha_B}-\ket{\alpha_{A}}\bra{\alpha_{B}}\otimes \mathbf{1}+\ket{\alpha_B}\hat \beta^\dagger_A)^{\tau_R}\\&+(\ket{\alpha_B}\hat (\beta^\dagger_A)^{\tau_R}-\ket{\alpha_{B}}\bra{\alpha_{A}}\otimes \mathbf{1})^{\tau_L}+A\leftrightarrow B,
    \end{split}
\end{align}
 where subscript $t$ stands for the total state and indices $A$ and $B$  for the corresponding parties. Note that swap operations $\tau_L,\tau_R$ can be performed using $N^2\times N^2$-dimensionl matrix $\hat \tau$, where $N$ is total number of modes:
 \begin{align} \label{FiB}
   \tau_{i,j}=
    \begin{cases}
        1 &  \exists m,n\in\{1...,d\}:\,\, i=m + d (-1 + n) \,\,\text{and} \,\,j =d (-1 + m) + n \\   
       0 & \text{otherwise}
    \end{cases}.
\end{align}
Then, for some operator $\hat O$, one gets:
\begin{align}
    O^{\tau_L}&=\hat\tau\hat O,\\
    O^{\tau_R}&=\hat O\hat\tau.
\end{align}

\section{Full second-order evolution}\label{app_second}
In a manner similar to the presented extended RSF approach, one could consider the full second-order evolution of the system. In such a scenario, one could consider reduction of the second-order observables to three reduced operators on single particle space:
\begin{align}
    \begin{split}
        \hat O&:=\sum_{k,k'}o_{k,k'}\hat{a}_k^\dagger\hat{a}_{k'}+\frac{1}{2}\sum_{k,k'}(q_{k,k'}\hat{a}_k\hat{a}_{k'}+q_{k,k'}^*\hat{a}_k^\dagger\hat{a}_{k'}^\dagger)\\
        &\rightarrow \Big(\hat{o}=\sum_{k,k'}o_{k,k'}\ket{k}\bra{k'},\hat{q}:=\frac{1}{2}\sum_{k,k'}(q_{k,k'})\ket{k}\bra{k'},\hat{q^*}:=\frac{1}{2}\sum_{k,k'}(q_{k,k'}^*)\ket{k}\bra{k'}\Big).\label{op_red}
        \end{split}
\end{align}
Note that reducing also first-order observables:
\begin{equation}
    \hat O'=\sum_k (p_k \hat a_k+p_k^*\hat a_k^\dagger)\rightarrow(\bra{p}:=\sum_k p_k\bra{k},\bra{p^*}:=\sum_k p_k^*\bra{k}),
\end{equation}
the expectation value of the combination of such observables is given by:
\begin{equation}
    \langle\hat O\rangle+\langle\hat O'\rangle=\Tr{\hat O\hat\rho_F}=\tr{(\hat o,\hat q,\hat q^*)\cdot(\hat\rho,\hat r,\hat r^*)^T}+(\bra{p},\bra{p^*})\cdot(\ket{\alpha},\ket{\alpha^*})^T.
\end{equation}
We can now allow any Hamiltonian that can be reduced by the procedure (\ref{op_red}). We denote the reduced Hamiltonian by the following triple $(\hat h, \hat h_s,\hat h_s^*)$. The operator $\hat h$ governs the free evolution and exchange of photons between modes, whereas the operators $\hat h_s,\hat h_s^*$ correspond to single-mode and two-mode squeezing. Introducing such an interaction Hamiltonian can correspond, e.g., to using a parametric approximation to some modes in the system. These modes are treated as classical coherent pumping fields which are subject to, e.g., a parametric down-conversion process in which annihilation of a single photon from the pumping field is accompanied by creation of two photons in modes which we treat in a full quantum manner. This addition to the formalism is similar to the addition of coherent pumping in (\ref{eq:full_ev}). However, such an extended evolution requires an extension of the reduced states. If one is only interested in the evolution of the original reduced state $\hat\rho$ the addition of $\hat r$ is sufficient. Then, one can analogously obtain equations of motion for RSF, which are the following:
\begin{align}
    \begin{split}
        \frac{d}{d t}\hat\rho&=-\frac{i}{\hbar} [\hat h,\hat\rho]-\frac{2i}{\hbar}(\hat h_s^*\hat r^*-\hat r\hat h_s)+(\ket{\xi}\bra{\alpha}+\ket{\alpha}\bra{\xi})+\frac{1}{2}\lbrace\hat \gamma_\uparrow-\hat \gamma_\downarrow^T ,\hat\rho \rbrace+\hat \gamma_\uparrow,\label{eq:ev_rho_ap}
    \end{split}\\
    \begin{split}
        \frac{d}{d t}\hat r&=-\frac{i}{\hbar} \hat h \hat r-\frac{2i}{\hbar}\hat h_s^{*}\hat \rho^T-\frac{i}{\hbar}\hat h_s^{*}+\ket{\xi}\bra{\alpha}+\frac{1}{2}(\hat \gamma_\uparrow-\hat \gamma_\downarrow^T)\hat r+ T.,\label{eq:ev_r_ap}
    \end{split}\\
        \begin{split}
        \frac{d}{d t}\ket{\alpha}&=-\frac{i}{\hbar} \hat h\ket{\alpha} -\frac{2i}{\hbar}\hat h_s^*\ket{\alpha}+\ket{\xi}+\frac{1}{2}(\hat \gamma_\uparrow-\hat \gamma_\downarrow^T)\ket{\alpha},\label{eq:ev_alpha_ap}
    \end{split}
\end{align}
where ``$T.$'' stands for the transposed expression. To consider the evolution of $\hat \rho_4$, one needs to extend the reduction by the additional reduced states:

\begin{align}
       \hat m&:=\sum_{k_1,k_2,k_3,k_4}\tr{\hat\rho_F\hat a_{k_1}^\dagger\hat a_{k_2}\hat a_{k_3}\hat a_{k_4}}\ket{k_2,k_4}\bra{k_1,k_3},\\
       \hat q&:=\sum_{k_1,k_2,k_3,k_4}\tr{\hat\rho_F\hat a_{k_1}\hat a_{k_2}\hat a_{k_3}\hat a_{k_4}}\ket{k_2,k_4}\bra{k_1,k_3},\\
              \hat \zeta&:=\sum_{k_1,k_2,k_3,k_4}\tr{\hat\rho_F\hat a_{k_1}\hat a_{k_2}\hat a_{k_3}}\ket{k_2,k_3}\bra{k_1}.
\end{align}
Then, the set of evolution equations is extended by:
\begin{align}
    \begin{split}
        \frac{d}{d t}\hat\rho_4&=-\frac{i}{\hbar} [\hat h\otimes \mathbf{1}+ \mathbf{1}\otimes \hat h ,\hat\rho_4]-\frac{2i}{\hbar}( \mathbf{1}\otimes \hat h_s^\dagger\hat m^\dagger+(\mathbf{1}\otimes \hat h_s^\dagger\hat m^\dagger)^{\tau_L}-\hat m\mathbf{1}\otimes \hat h_s-(\hat m \mathbf{1}\otimes \hat h_s)^{\tau_R})\\&-\frac{2i}{\hbar}\left((\hat h_s^\dagger\hat r^\dagger\otimes\mathbf{1})^{\tau_L}+\left[(\hat r^\dagger\otimes\hat h_s^\dagger)^{\tau_L}\right]^{T_2}-(\hat r\hat h_s\otimes\mathbf{1})^{\tau_L}-\left[(\hat h_s\otimes \hat r)^{\tau_L}\right]^{T_2}\right)\\
        &+\left(\ket{\xi}\hat\beta^\dagger+(\ket{\xi}\hat\beta^\dagger)^{\tau_L}+(\ket{\xi}\bra{\alpha}\otimes\mathbf{1})^{\tau_L}+\hat\beta\bra{\xi}+(\hat\beta\bra{\xi})^{\tau_R}+(\mathbf{1}\otimes\ket{\alpha}\bra{\xi})^{\tau_R}\right)\\
        &+\frac{1}{2}\lbrace(\mathbf{1}\otimes\hat\gamma_\uparrow+\hat\gamma_\uparrow\otimes\mathbf{1})- (\mathbf{1}\otimes\hat\gamma_\downarrow^T+\hat\gamma_\downarrow^T\otimes\mathbf{1}) ,\hat\rho \rbrace\\&+(\hat\rho\otimes\gamma_\downarrow^T)^{\tau_L}+(\gamma_\uparrow\otimes\hat\rho)^{\tau_L}+\gamma_\uparrow\otimes\hat\rho+\hat\rho\otimes\gamma_\uparrow+(\gamma_\uparrow\otimes\mathbf{1})^{\tau_L},
    \end{split}\\
    \begin{split}
        \frac{d}{d t}\hat m&=-\frac{i}{\hbar}[\hat h \otimes \mathbf{1}, \hat m ]-\frac{i}{\hbar} (\mathbf{1}\otimes\hat h \hat m+\hat m\mathbf{1}\otimes\hat h^T)-\frac{2i}{\hbar}(\hat \rho_4 \mathbf{1}\otimes\hat h_s^\dagger+\lbrace\mathbf{1}\otimes\hat h_s^\dagger,\hat \rho_4\rbrace^{T_2}-\hat q \hat h_s\otimes\mathbf{1})\\
        &-\frac{2i}{\hbar}\left(\hat\rho \otimes h_s^\dagger-\left[(\hat\rho \otimes h_s^\dagger)^{\tau_L}\right]^{T_2}\right)+\left(\hat\beta\bra{\xi^*}+(\hat\beta\bra{\xi^*})^{T_2}+\left[(\hat\beta\bra{\xi^*})^{T_2}\right]^{\tau_L}-(\hat\zeta \bra{\xi})^{\tau_R}\right)\\
        &+\frac{1}{2}\lbrace(\hat\gamma_\uparrow\otimes\mathbf{1})- (\hat\gamma_\downarrow^T\otimes\mathbf{1}) ,\hat m \rbrace+\frac{1}{2}(\mathbf{1}\otimes\hat\gamma_\uparrow \hat m+\hat m \mathbf{1}\otimes\hat\gamma_\uparrow^T-  \mathbf{1}\otimes\hat\gamma_\downarrow^T\hat m-\hat m \mathbf{1}\otimes\hat\gamma_\downarrow)\\
        &-\hat\gamma_\uparrow\otimes\hat r-(\hat\gamma_\uparrow\otimes\hat r)^{\tau_L}-\left[(\hat\gamma_\uparrow\otimes\hat r)^{\tau_L}\right]^{T_2},
    \end{split}\\
        \begin{split}
        \frac{d}{d t}\hat q&=-\frac{i}{\hbar} (\hat h \otimes\mathbf{1} \hat q+\hat q\hat h^T \otimes \mathbf{1}+\mathbf{1}\otimes\hat h \hat q+\hat q\mathbf{1}\otimes\hat h^T)\\&-\frac{2i}{\hbar}\left((\hat h_s^\dagger\otimes\mathbf{1} \hat m^T)^{\tau_L}+( \hat m\hat h_s^\dagger\otimes\mathbf{1})^{\tau_R}+ \hat h_s^\dagger\otimes\mathbf{1} \hat m+ \hat m^T\hat h_s^\dagger\otimes\mathbf{1} \right)\\
        &-\frac{2i}{\hbar}\left( \hat r\otimes\hat h_s^\dagger+ (\hat r\otimes\hat h_s^\dagger)^{\tau_L}+ \left[(\hat r\otimes\hat h_s^\dagger)^{\tau_L}\right]^{T_2}\right)-\frac{i}{\hbar}\left( \hat h_s^\dagger\otimes \hat r+ (\hat h_s^\dagger\otimes \hat r)^{\tau_L}+ \left[(\hat h_s^\dagger\otimes \hat r)^{\tau_L}\right]^{T_2} \right)\\
        &+\left(\hat \zeta \bra{\xi^*}+(\hat \zeta \bra{\xi^*})^{T_2}+(\hat \zeta \bra{\xi^*})^{\tau_R}+\left[ (\hat \zeta \bra{\xi^*})^{\tau_R}\right]^{T_1}\right)\\
        &+\frac{1}{2}(\hat \gamma_\uparrow\otimes\mathbf{1}+\mathbf{1}\otimes \hat \gamma_\uparrow-\hat \gamma_\downarrow^T\otimes\mathbf{1} -\mathbf{1}\otimes\hat \gamma_\downarrow^T)\hat  q+\frac{1}{2}\hat q(\hat \gamma_\uparrow^T\otimes\mathbf{1}+\mathbf{1}\otimes \hat \gamma_\uparrow^T-\hat \gamma_\downarrow\otimes\mathbf{1} -\mathbf{1}\otimes\hat \gamma_\downarrow),
    \end{split}\\
       \begin{split}
        \frac{d}{d t}\hat \beta&=-\frac{i}{\hbar}[\hat h\otimes\mathbf{1} ,\hat \beta]-\frac{i}{\hbar}\mathbf{1}\otimes\hat h\hat \beta\\&-\frac{2i}{\hbar}\left( (\hat\beta^\dagger\mathbf{1}\otimes\hat h_s^\dagger)^{T_2}+\left[(\hat\beta^\dagger\mathbf{1}\otimes\hat h_s^\dagger)^{T_2}\right]^{\tau_L}+\left[(\hat h_s^\dagger\otimes \bra{\alpha})^{\tau_R}\right]^{T_2}-\hat \zeta \hat h_s\otimes\mathbf{1}\right)\\
        &+\left(\hat \rho\otimes\ket{\xi}+(\hat \rho\otimes\ket{\xi})^{\tau_L}-\left[(\hat r\otimes\bra{\xi})^{\tau_R}\right]^{T_2} \right)\\&+\frac{1}{2}(\lbrace\hat\gamma_\uparrow\otimes\mathbf{1}-\hat\gamma_\downarrow^T\otimes\mathbf{1},\hat \beta\rbrace+(\mathbf{1}\otimes\hat\gamma_\uparrow-\mathbf{1}\otimes\hat\gamma_\downarrow^T)\hat\beta)
        +\hat\gamma_\uparrow\otimes\ket{\alpha}+(\hat\gamma_\uparrow\otimes\ket{\alpha})^{\tau_L},
        \end{split}\\
        \begin{split}
        \frac{d}{d t}\hat \zeta&=-\frac{i}{\hbar}[\hat h\otimes\mathbf{1} \hat \zeta+\hat \zeta \hat h^T\otimes\mathbf{1}+\mathbf{1}\otimes\hat h\hat \zeta]-\frac{2i}{\hbar}\left(\hat \beta\hat h_s^\dagger\otimes\mathbf{1}+\hat h_s^\dagger\otimes\mathbf{1}\hat \beta^{T_1}+(\hat h_s^\dagger\otimes\mathbf{1}\hat \beta^{T_1})^{\tau_L} \right)\\
        &-\frac{2i}{\hbar}\left( \hat h_s^\dagger\otimes\ket{\alpha}+(\hat h_s^\dagger\otimes\ket{\alpha})^{\tau_L}+\left[(\hat h_s^\dagger\otimes\ket{\alpha})^{\tau_L}\right]^{T_1}\right)\\&+\left(\hat r\otimes\ket{\xi}+(\hat r\otimes\ket{\xi})^{\tau_L}+\left[(\hat r\otimes\ket{\xi})^{\tau_L}\right]^{T_1} \right)\\
        &+\frac{1}{2}\left((\hat\gamma_\uparrow\otimes\mathbf{1}-\hat\gamma_\downarrow^T\otimes\mathbf{1})\hat \zeta+\hat\zeta (\hat\gamma_\uparrow^T\otimes\mathbf{1}-\hat\gamma_\downarrow\otimes\mathbf{1})+(\mathbf{1}\otimes\hat\gamma_\uparrow-\mathbf{1}\otimes\hat\gamma_\downarrow^T)\hat\zeta\right).
        \end{split}
\end{align}
Note that such an extended RSF allows for calculation of any additive observable up to fourth the order in creation and annihilation operators.

\end{document}